\documentclass[aps,prd,reprint,twocolumn,superscriptaddress,preprintnumbers,nofootinbib]{revtex4-2}
\usepackage{graphicx,color,xcolor}
\usepackage{amsmath,amssymb,amsfonts,amsthm}
\usepackage{bbm}
\usepackage[caption=false]{subfig}
\usepackage[colorlinks=True, citecolor=blue, urlcolor=blue, linkcolor=blue]{hyperref}
\usepackage{orcidlink}
\usepackage{qcircuit}
\usepackage{soul}
\usepackage{wasysym}
\usepackage{physics}
\usepackage{mathrsfs}
\usepackage{enumitem}

\newcommand{\nn}{\nonumber}

\newcommand{\ml}{\mathcal}
\newcommand{\bs}{\boldsymbol}

\newcommand{\ba}{\begin{aligned}}
\newcommand{\ea}{\end{aligned}}
\newcommand{\eq}[1]{Eq.~\eqref{eq:#1}}
\definecolor{darkgreen}{rgb}{0.13,0.55,0.13}

\begin{document}
\title{Quantum Computing for Energy Correlators}

\author{Kyle Lee \orcidlink{0000-0002-1827-2958}}
\email{kylel@mit.edu}
\affiliation{Center for Theoretical Physics, Massachusetts Institute of Technology, Cambridge, MA 02139, USA}

\author{Francesco Turro \orcidlink{0000-0002-1107-2873}}
\email{francesco.turro@gmail.com}
\affiliation{InQubator for Quantum Simulation, University of Washington, Seattle, WA 98195, USA}

\author{Xiaojun Yao \orcidlink{0000-0002-8377-2203}}
\email{xjyao@uw.edu}
\affiliation{InQubator for Quantum Simulation, University of Washington, Seattle, WA 98195, USA}

\date{\today}
\preprint{MIT-CTP 5767, IQuS@UW-21-089}
\begin{abstract}
In recent years, energy correlators have emerged as powerful observables for probing the fragmentation dynamics of high-energy collisions. We introduce the first numerical strategy for calculating energy correlators using the Hamiltonian lattice approach, providing access to the intriguing nonperturbative dynamics of these observables. Furthermore, motivated by rapid advances in quantum computing hardware and algorithms, we propose a quantum algorithm for calculating energy correlators in quantum field theories. This algorithm includes ground state preparation, the application of source, sink, energy flux and real-time evolution operators, and the Hadamard test. We validate our approach by applying it to the SU(2) pure gauge theory in $2+1$ dimensions on $3\times 3$ and $5\times 5$ honeycomb lattices with $j_{\rm max} = \frac{1}{2}$ at various couplings, utilizing both classical methods and the quantum algorithm, the latter tested using the IBM emulator for specific configurations. The results are consistent with the expected behavior of the strong coupling regime and motivate a more comprehensive study to probe the confinement dynamics across the weak and strong coupling regimes.
\end{abstract}

\maketitle

\section{Introduction}
\begin{figure}
\includegraphics[width=0.25\textwidth]{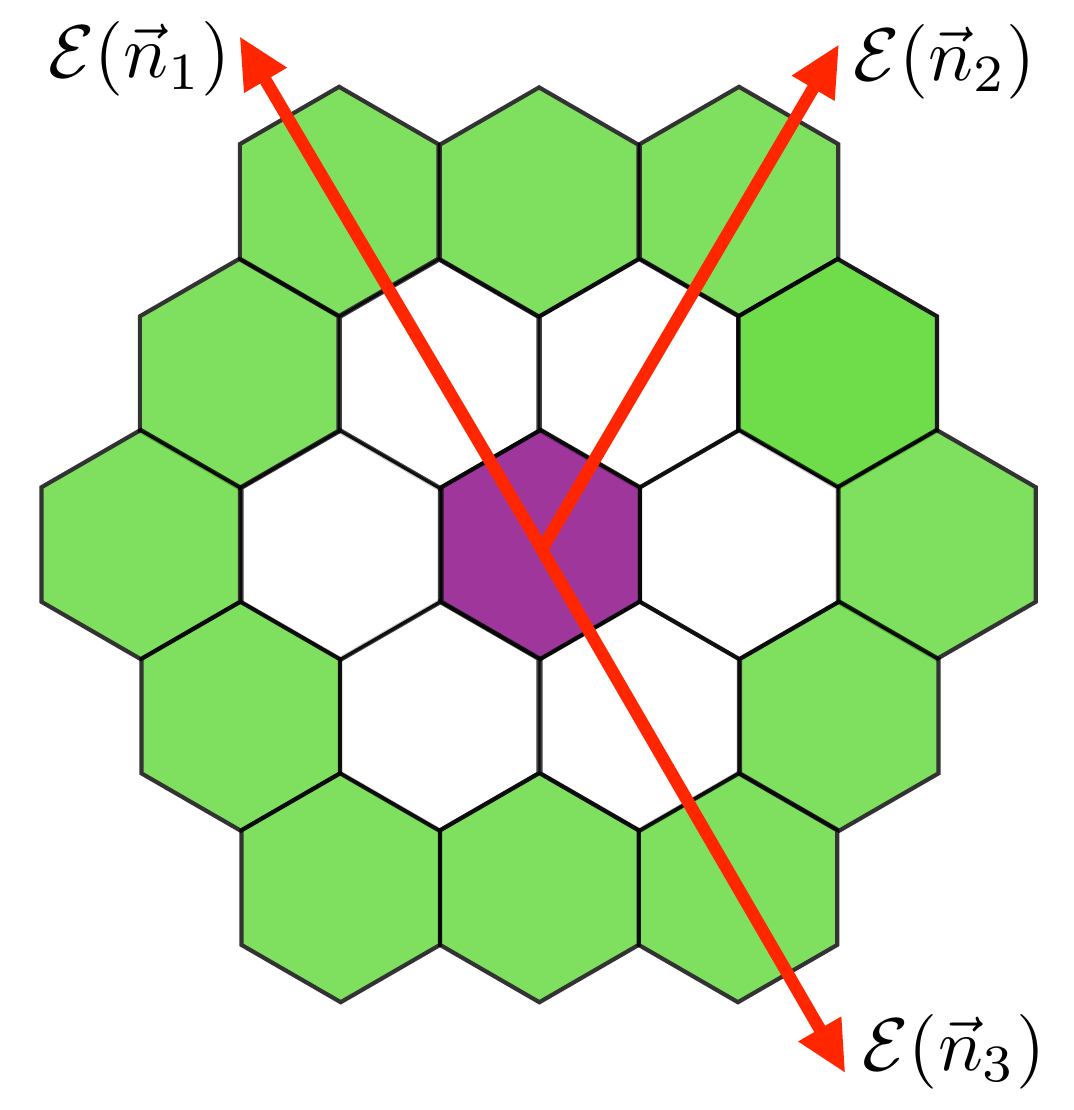}
\caption{A schematic diagram of energy flows on a 2-dimensional honeycomb lattice with the source created at the center, marked in purple.}
\label{fig:scheme}
\end{figure}

Energy correlators are among the most natural Lorentzian field theory observables with many applications in both conformal field theories (CFTs) and high-energy collider experiments. They are defined by the correlation functions of the form $\langle \mathcal{E}(\vec{n}_1) \mathcal{E}(\vec{n}_2)\cdots  \mathcal{E}(\vec{n}_N)\rangle\equiv \langle \Psi | \mathcal{E}(\vec{n}_1) \mathcal{E}(\vec{n}_2)\cdots  \mathcal{E}(\vec{n}_N)|\Psi\rangle$ where the energy flow operator in $d+1$ dimensions~\cite{Basham:1979gh,Basham:1978zq,Basham:1978bw}
\begin{align}
\label{eq:EF}
\mathcal{E}(\vec{n}) = \lim_{r \rightarrow \infty} \int_0^\infty {\rm d}t\, r^{d-1}\, n^i T_{0 i}(t, r \vec{n})\,,
\end{align}
encapsulates the asymptotic energy flux $T_{0i}$ on the celestial sphere in the direction $\vec{n}$ from a normalized 
state $|\Psi \rangle = \frac{\mathcal{O}|0\rangle}{\sqrt{\langle 0| \ml{O}^\dagger \ml{O} |0\rangle}}$ 
created by a local operator $\mathcal{O}$ as in Fig.~\ref{fig:scheme}. 

The energy flow operator in \eq{EF} and its correlations have been widely studied in CFTs~\cite{Chicherin:2024ifn,He:2024hbb,Korchemsky:2015ssa,Belitsky:2013ofa,Korchemsky:2019nzm,Caron-Huot:2022eqs,Belin:2020lsr,Korchemsky:2021htm,Kravchuk:2018htv,Cordova:2018ygx,Balakrishnan:2019gxl,Firat:2023lbp,Chicherin:2023gxt,Korchemsky:2021htm,Besken:2020snx,Cordova:2017zej,Cordova:2017dhq,Belitsky:2013bja}. Notably, this has led to an improved understanding of the operator product expansion (OPE) for Lorentzian operators, specifically the light-ray OPE~\cite{Chang:2020qpj,Kologlu:2019mfz,Hofman:2008ar}.
In recent years, there has been a strong motivation to measure such light-ray OPE in collider experiments by studying the substructure of jets~\cite{Larkoski:2017jix,Marzani:2019hun}, where the energy flow operator in \eq{EF} represents the experimental detectors that capture the energy flux. (For this reason, we will use `energy flow operator' and `detector' interchangeably.) This has resulted in the observation of universal scaling predicted by the light-ray OPE across multiple experimental collaborations~\cite{Fan2023,Tamis:2023guc,CMS:2024mlf}, along with a surge in phenomenological studies using energy correlators to explore the Lorentzian dynamics of quantum chromodynamics (QCD)~\cite{Schindler:2023cww,Bossi:2024qho,Holguin:2024tkz,Holguin:2023bjf,Xiao:2024rol,Holguin:2022epo,Chen:2023zlx,Lee:2022ige,Chen:2020vvp,Jaarsma:2022kdd,Jaarsma:2023ell,Chen:2022muj,Lee:2023npz,Lee:2023tkr,Lee:2024esz,Chen:2024nyc,Andres:2023xwr,Andres:2022ovj,Devereaux:2023vjz,Andres:2023ymw,Andres:2024ksi,Barata:2024nqo,Barata:2023bhh,Cao:2023oef,Chen:2024bpj,Liu:2023aqb,Liu:2022wop,Craft:2022kdo,Chen:2020adz,Yang:2024gcn,Chang:2022ryc,Chen:2019bpb,Andres:2024hdd,Singh:2024vwb}. 

One of the most interesting features of energy correlators is that they exhibit qualitatively different behaviors in the weak and strong coupling regions. In the weak coupling regime, the light-ray OPE dictates a power-law angular scaling of correlators, determined by the anomalous dimensions of twist-2 operators. 
In contrast, rapid fragmentation in the strong coupling region~\cite{Polchinski:2002jw,Strassler:2008bv} produces a large number of uncorrelated soft particles, causing the leading behavior of $N$-point correlators to become angle-independent and reduce to a product of one-point correlators as%
\footnote{A large number of quanta created from heavy or large charge operators also exhibit such homogeneous distribution as the leading behavior, regardless of the value of the coupling, as demonstrated in Refs.~\cite{Firat:2023lbp,Chicherin:2023gxt}.}
\begin{align}
\label{eq:strongcoupl}
\langle \mathcal{E}(\vec{n}_1) \mathcal{E}(\vec{n}_2)\cdots  \mathcal{E}(\vec{n}_N)\rangle \approx  \langle \mathcal{E}\rangle^N = \left(\frac{Q}{\Omega_{d}}\right)^N\,,
\end{align}
where $Q$ is the total energy flow and $\Omega_d = \frac{2\pi^{d/2}}{\Gamma(d/2)}$ is the area of the celestial sphere for $d+1$-dimensional theories.

These qualitatively distinct behaviors in the weak and strong coupling regions manifest in the most interesting ways in QCD. Small-angle correlations inside high-energy jets exhibit the OPE scaling dictated by the twist-2 operators of the underlying partons as the angle $\theta_{ij} = \arccos (\vec{n}_i \cdot \vec{n_j})$ is reduced, until the relative transverse energy in the branching becomes nonperturbative, $Q\theta_{ij} \sim \Lambda_{\rm QCD}$, resulting in a dense cloud of bounded hadrons that are uncorrelated. This remarkable transition from the parton to hadron regime~\cite{Lee:2022ige,Lee:2024esz,Komiske:2022enw} has now been observed in experimental data~\cite{Fan2023,Tamis:2023guc,CMS:2024mlf} and is of utmost interest for advancing our understanding of confinement. However, while the correlators in the strong coupling region are calculable for gauge theories with gravitational duals~\cite{Hofman:2008ar,Chen:2024iuv}, analytical methods are not currently feasible for general confining gauge theories like QCD.

In recent years, the Hamiltonian lattice approach emerged as a promising tool for nonperturbative studies of the Lorentzian dynamics of quantum field theories (QFTs), driven by advances in quantum computing hardware and algorithms, as well as classical techniques for real-time dynamics~\cite{Bender:2020ztu,Warrington:2023aqa,Klco:2018kyo,banuls2020simulating,klco2022standard,Davoudi:2022xmb,Bauer_2023,beck2023quantum,bauer2023quantum,Lamm:2019bik,DeJong:2020riy,Bauer:2021gek,DAndrea:2023qnr,Halimeh:2023lid,miao2023quantum,magnifico2024tensor,miao2023convergence,Fontana:2024rux,Kadam:2022ipf,Watson:2023oov,Kadam:2024zkj,Li:2024lrl,Kavaki:2024ijd,Ciavarella:2024fzw,Araz:2024bgg,Araz:2024xkw,Davoudi:2024osg,Mueller:2024mmk,trivedi2024quantum,kashyap2024accuracy,Grabowska:2024emw}. Recent progress includes the development of classical and quantum algorithms for simulating gauge theories, which are central to our understanding of high-energy physics. Notably, there have been successful demonstrations of classical and quantum simulations of lattice gauge theories in lower dimensions, providing a foundation for exploring more intricate theories relevant to QCD~\cite{Lee:2023urk,Lin:2024eiz,Farrell:2024mgu,Illa:2024kmf,Davoudi:2024wyv,Farrell:2024fit,Gustafson:2024bww,Calajo:2024bvs,Desaules:2024cua,Desaules:2024cua,Kebric:2023nnd,Guo:2024tnb,Angelides:2023noe,Angelides:2023bme,Belyansky:2023rgh,Cataldi:2023xki,deJong:2021wsd,Florio:2023dke,Florio:2024aix,Ebner:2024mee,Ebner:2023ixq}.

In this paper, we capitalize on these advancements and develop a quantum algorithm to compute energy correlators nonperturbatively for a generic QFT using the Hadamard test~\cite{Lin:2022vrd}. As a demonstration, we consider the $2+1$-dimensional SU(2) pure gauge theory, a confining theory~\cite{Karabali:1996je,Karabali:1997wk,Polyakov:1976fu,tHooft:1977nqb,Jackiw:1980kv,Deser:1982vy}, and calculate energy correlators on a small lattice with fixed couplings and truncated local Hilbert spaces. We test the quantum algorithm on a $3\times 3$ honeycomb lattice and classically calculate energy correlators on both $3\times 3$ and $5 \times 5$ honeycomb lattices as schematically shown in Fig.~\ref{fig:scheme}. Our study lays the groundwork for future calculations in the continuum limit and demonstrates how quantum computing can be valuable for high-energy physics, even without the complicated initial wave packet preparation typically required for scattering simulations~\cite{Jordan:2011ci,Jordan:2012xnu}, although progress has been made to prepare wave packets~\cite{Farrell:2024fit,Davoudi:2024wyv}.

The paper is organized as follows: In Sec.~\ref{sec:algorithm}, we will discuss the general methods for computing energy correlators using the Hamiltonian lattice approach, followed by a description of the classical and quantum algorithms employed. In Sec.~\ref{sec:su2}, we will review the $2+1$-dimensional SU(2) Hamiltonian lattice gauge theory on a honeycomb lattice and outline the construction of the relevant operators for energy correlator computations. In Sec.~\ref{sec:results}, our results of the energy correlator will be presented for the truncated $2+1$-dimensional SU(2) gauge theory on the honeycomb lattice, computed using both classical and quantum algorithms. We will conclude in Sec.~\ref{sec:conclusions}.

\section{Calculating Energy Correlators in the Hamiltonian Lattice approach}
\label{sec:algorithm}
In this section, we first outline the general numerical method for computing energy correlators of $d+1$-dimensional QFTs via the Hamiltonian lattice approach. This includes explaining the construction of multiple energy flow operators by applying detector limits on the lattice. We then describe both the classical and quantum algorithms that will be used. For simplicity, we primarily focus on the two-point energy correlator, though the discussion can be readily extended to higher-point correlators.

\subsection{Energy correlators from detector limits}
\label{subsec:detector}

The operator definition of the energy correlators $\langle \ml{E}({\vec n}_1) \ml{E}({\vec n}_2) \rangle$ involves the non-time-ordered Wightman correlation function%
\footnote{
Local Lorentzian correlations, or Wightman functions, are defined with time ordering in Euclidean imaginary time to suppress unbounded energy states~\cite{Belin:2020lsr,Firat:2023lbp,Hartman:2015lfa}. In the lattice framework, the finite lattice size $a$ naturally regulates high energy states, preventing divergence.} of local Lorentzian operators $\langle 0| \ml{O}^\dagger n_1^i T_{0i}(t_1,r_1 \vec{n}_1)n_2^jT_{0j}(t_2,r_2 \vec{n}_2)\ml{O} |0\rangle$, where $\ml{O}$ represents either a localized or an extended\footnote{Extended source that can be written as an integration of localized source.} source operator generating a perturbation on top of the vacuum, and $\ml{O}^\dagger$ is the corresponding sink operator.

\subsubsection{State created by local source}
\begin{figure}[t]
\includegraphics[width=0.308\textwidth]{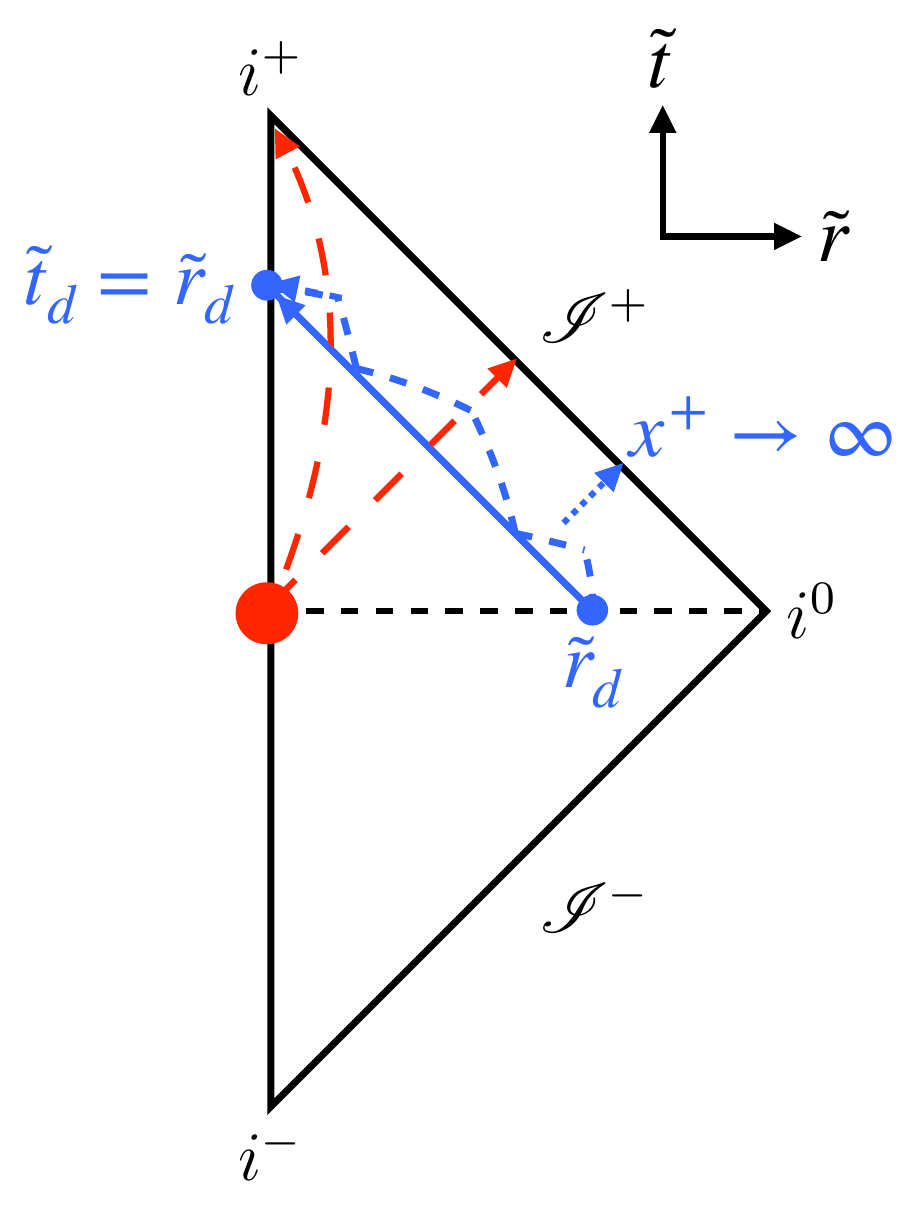}
\caption{Penrose diagram for $d+1$ dimensional Minkowski space depicting the integration and limiting procedure for constructing an energy flow operator. Each point represents $d-1$ dimensional sphere. The Penrose coordinates $\tilde{t}$ and $\tilde{r}$ are related to the Minkowski coordinates $t$ and $r$ as $\tan(\tilde{t}\pm\tilde{r}) = t\pm r = x^\pm$, and manifest the causal structure of the spacetime. The red point represents a local source of perturbation from which the signals travel. The two red dashed lines indicate different paths taken by massive and massless excitations from the source at $(t,r)=(0,0)$, reaching timelike infinity $i^+$ and null future infinity $\mathscr{I}^+$, respectively. The $i^0$ indicates spacelike infinity. The blue solid line represents the continuum integration path of the detector that is turned on at an initial position $r_d$ from $t=0$ to $t_d$ with $x^+$ constant. The blue dashed curved line approximates this path for lattice calculations. We take the $x^+\to\infty$ limit by taking $r_d$ to $i^0$ and $t_d$ to $i^+$ while moving the blue solid line along the blue dotted arrow line.}
\label{fig:penrose1}
\end{figure}

We first consider the case where both $\ml{O}$ and $\ml{O}^\dagger$ are local operators inserted at the same spacetime point $x$, i.e., $\ml{O}=\ml{O}(x) = \ml{O}(t,\vec{x})$, $\ml{O}^\dagger=\ml{O}^\dagger(x)=\ml{O}^\dagger(t,\vec{x})$. Without loss of generality, we assume they are placed at the spacetime origin $x=0$, denoted by the red point in Fig.~\ref{fig:penrose1}. To compute the energy correlators, one needs to employ a detector limiting procedure as in~\eq{EF}, i.e. light transform~\cite{Hofman:2008ar,Kravchuk:2018htv,Belitsky:2013xxa,Chicherin:2020azt}, by sending the stress-energy tensor $T_{0i}$ to infinity and performing the time integration. This procedure represents constructing an asymptotic detector which turn on during some detector operation time.

\begin{figure}[t]
\includegraphics[width=0.3\textwidth]{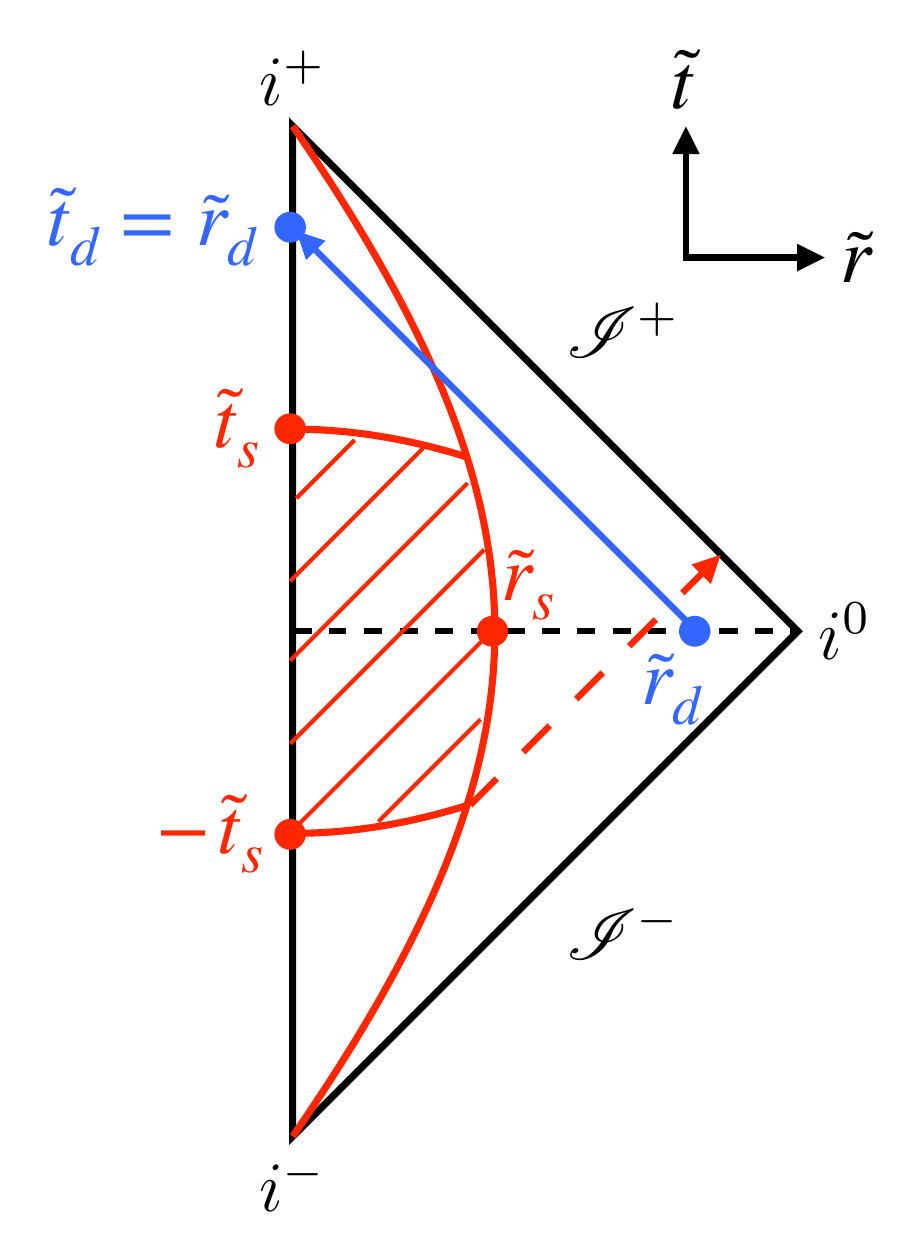}
\caption{Same as in Fig.~\ref{fig:penrose1} except for an extended source in spacetime, indicated by the red shaded region defined by $t_s$ and $r_s$. By placing the detector at $r_d\geq t_s + r_s$ first and turning it on at $t=0$, no signals will be missed.}
\label{fig:penrose2}
\end{figure}

The precise order between such a limit and the integral in \eq{EF} is ambiguous. Following Refs.~\cite{Belitsky:2013xxa,Hofman:2008ar}, we adopt a procedure of integrating over the retarded time $x^-\equiv t-r$ while keeping the advanced time $x^+\equiv t+r$ fixed, as depicted by the blue solid line in Fig.~\ref{fig:penrose1}, and sending $x^+$ to infinity, as indicated by the blue dotted line. This procedure can be written as
\begin{align}
\label{eq:EE_limit1}
&\langle 0| \ml{O}^\dagger(0) \ml{E}({\vec n}_1) \ml{E}({\vec n}_2) \ml{O}(0) |0\rangle = \lim_{x^+\to\infty}(x^{+})^{2d-2} \, G_{{\vec n}_1{\vec n}_2}(x^+) \,,\nn\\
&G_{{\vec n}_1{\vec n}_2}(x^+) = \frac{1}{4} \int_{-x^+}^{x^+}{\rm d}x^-_1 \int_{-x^+}^{x^+}{\rm d}x^-_2 \nn\\
& \langle 0| \ml{O}^\dagger(0) n^i_1T_{-i}(x_1^-,x^+,\vec{n}_1) n^j_2T_{-j}(x_2^-,x^+,\vec{n}_2) \ml{O}(0) | 0\rangle \,,
\end{align}
where the integrated correlation $G_{{\vec n}_1{\vec n}_2}(x^+)$ is at a fixed advanced time. We have used the fact that the QFT is in $d+1$ dimensions. We also note that the bound for $x^-$ is determined assuming that the source is inserted at $t=0$. This procedure is natural for massless excitations, such as those in CFTs ending in the future null infinity. For gapped theories like the SU(2) gauge theory we consider later, the integration over the retarded time should be taken before sending the advanced time to infinity, in order to capture massive excitations at timelike infinity, as noted in Ref.~\cite{Csaki:2024joe}. 

Keeping all the detectors at fixed $x^+$ while the integration is performed maintains spacelike separation between the detectors on the same light-sheet even for finite $r$. For $\vec{n}_1\neq \vec{n}_2$, which we focus on in this paper, we then expect the energy flow operators to commute~\cite{Kravchuk:2018htv,Korchemsky:2021htm,Belin:2020lsr,Besken:2020snx}
\begin{align}
\label{eq:commutation}
[\ml{E}({\vec n}_1), \ml{E}({\vec n}_2)] = 0\,.
\end{align}

Due to the discretization inherent in the lattice, we must also discretize our detector limiting procedure. We approximate the smooth integration path with a discretized version, as shown by the blue dashed line in the Penrose diagram in Fig.~\ref{fig:penrose1}. Following this numerical procedure, we can approximate $G_{{\vec n}_1{\vec n}_2}(x^+)$ in Eq.~\eqref{eq:EE_limit1} on the lattice as
\begin{align}
\label{eq:lattice_time_inte}
& G_{{\vec n}_1{\vec n}_2}(x^+)\big|_{\rm lattice} = \sum_{i=0}^{N} \int_{t_{i}}^{\tilde{t}_{i}}{\rm d}t_1\int_{t_{i}}^{\tilde{t}_{i}} {\rm d}t_2 \nn\\
& \qquad \langle 0| \ml{O}^\dagger(0) n^i_1T_{0i}(t_1,r_{i}{\vec n}_1) n^j_2T_{0j}(t_2,r_{i}{\vec n}_2) \ml{O}(0) |0\rangle \,.
\end{align}
Here, $i$ in the sum denotes the lattice integration step with time between $t_{i}$ and $t_{i+1}$ at $r_{i}$. Explicitly, we take
\begin{align}
t_{i}&= t_{0}+i\times\frac{(t_{N}-t_{0})}{N} \,, \quad t_i<\tilde{t}_{i}\leq t_{i+1}\,,\nn\\
r_{i}&= r_{0}+i\times\frac{(r_{N}-r_{0})}{N} \,,
\end{align}
with the boundary values
\begin{align}
t_{0}&=0\,,\quad &t_{N}&=x^+\nn\\
r_{0}&=x^+\,,\quad  &r_{N}&=0\,,
\end{align}
where $x^+$ corresponds to $r_d$ in Fig.~\ref{fig:penrose1}.

We note that at each step $i$, we fix $r_{i}+t_{i}=x^+$ as desired. We also take $t_{0}=0$ as we identify $t=0$ with the time we insert the source. In the Hamiltonian lattice theory setup, time is continuous but space is discretized. The smallest step we can take is given by the lattice spacing $a$, i.e., $r_0-r_N=Na$. (To reduce the lattice cutoff artifact, one should take each step to be a few lattice spacings, say $M a$, which would give $r_0-r_N = N M a$.) At finite $a$, the relation to keep $x^+$ fixed is slightly broken when we integrate between $t_{i}$ and ${t}_{i+1}$ with fixed $r_{i}$. This can spoil the spacelike separation that we want to ensure in \eq{commutation}, leading to causal interference between detectors. To mitigate this undesired lattice artifact, we set the upper limit of the time integration at each step $i$ to $\tilde{t}_i \leq t_{i+1}$, which means we miss energy flux between $\tilde{t}_i$ and $t_{i+1}$. However, as the lattice spacing $a \to 0$, these artifacts should disappear, allowing us to take $\tilde{t}_i \to t_{i+1}$ in the continuum limit.
We will explain our choice of $\tilde{t}_i$ later in our explicit setup for the $2+1$ SU(2) theory on the honeycomb lattice.

\subsubsection{Momentum eigenstate}
In the previous section, we discussed local sources and assumed that both $\mathcal{O}$ and $\mathcal{O}^\dagger$ were inserted at the origin of spacetime. In this subsection, we introduce a strategy for calculating energy correlators of momentum eigenstates that are not localized. For example, in $e^+e^-$ collisions, the electromagnetic current $J^\mu(x)=e\bar{\psi}(x) \gamma^\mu \psi(x)$ can source a state with definite energy and momentum $q$, then we can define an energy-energy correlator as (see e.g.~Ref.~\cite{Belitsky:2013xxa}) 
\begin{align}
\label{eq:EECee}
\langle &\ml{E}({\vec n}_1) \ml{E}({\vec n}_2) \rangle_q \nn\\
&\equiv \frac{1}{\sigma_{\rm tot}} \int{\rm d}^4x\,e^{iq\cdot x} \langle 0| J_\mu^\dagger(x) \ml{E}({\vec n}_1) \ml{E}({\vec n}_2) J^\mu(0) |0\rangle \,,
\end{align}
where the normalization factor is given by $\sigma_{\rm tot} \equiv \int{\rm d}^4x\, e^{iq\cdot x} \langle 0| J_\mu^\dagger(x) J^\mu(0) |0\rangle $.
In other words, general momentum eigenstate created in a high-energy scattering process involves source and sink term that spans the whole spacetime rather than a single fixed spacetime point as considered thus far. To implement this numerically on a lattice, we can adopt a limiting procedure by introducing a Gaussian suppression factor given by
\begin{align}
\label{eq:gauss}
\langle &\ml{E}({\vec n}_1) \ml{E}({\vec n}_2) \rangle_q =\lim_{\sigma\to \infty}\frac{1}{\sigma_{\rm tot} } \int{\rm d}^4x\,e^{iq\cdot x}e^{-\frac{x_0^2+x_1^2+x_2^2+x_3^2}{\sigma^2}}\nn\\
&\hspace{2cm}\times\langle 0| J_\mu^\dagger(x) \ml{E}({\vec n}_1) \ml{E}({\vec n}_2) J^\mu(0) |0\rangle \,,
\end{align}
which is same as~\eq{EECee} in the limit. However, for a finite value of $\sigma$, the Gaussian suppression factor ensures that \eq{gauss} has finite support with a width of $\sigma$ around the point $(t, \vec{x}) = (0, \vec{0})$. Neglecting small contributions from the Gaussian tail, we can define a time extent $t_s$ and a spatial extent $r_s$, within which the source and sink terms are applied, as illustrated in Fig.~\ref{fig:penrose2}. Unlike the scenario in Fig.~\ref{fig:penrose1}, where the source and sink terms are inserted at a single point, the finite region necessitates placing the detector sufficiently far away to ensure that no signals are missed. Signals traveling the farthest are emitted radially outward from $(-t_s, r_s)$ and reach $r = t_s + r_s$ at $t = 0$. Consequently, the detector must be positioned at $r_d \geq t_s + r_s$ as in Fig.~\ref{fig:penrose2} to capture all signals from the source. In other words, the width $\sigma$ effectively defines a finite region where the source and sink terms are inserted, which then imposes a lower bound on the distance at which the detector must be placed. We can then follow the implementation in \eq{lattice_time_inte} with $r_0=x^+ \geq t_s+r_s$ and an additional integral over the finite region controlled by $\sigma$, where the source and sink terms are inserted, to study energy correlators of the momentum eigenstate in the limit of \eq{gauss}.

We find it particularly intriguing that, by this method, we can emulate scattering processes in high-energy collisions using quantum computing without the usual challenges of initial wave packet preparation, which often present significant hurdles and extra computing resources~\cite{Jordan:2011ci,Jordan:2012xnu}, although progress has been made to prepare wave packets~\cite{Farrell:2024fit,Davoudi:2024wyv}. Instead, we assume that an initial scattering event produces a definite momentum eigenstate by some operators, similar to what occurs in $e^+e^-$ collisions, and then focus on studying the signals that arise from this momentum eigenstate.

\subsection{Real-time correlator from Hamiltonian dynamics and classical simulation}
In the case of local source,
the real-time local Lorentzian correlations we need for energy correlators can be written in the Heisenberg picture as 
\begin{align}
\label{eq:time_correlation1}
& \langle 0| \ml{O}^\dagger(0,\vec{0})n^i_1T_{0i}(t_1,r{\vec n}_1) n^j_2T_{0j}(t_2,r{\vec n}_2) \ml{O}(0,\vec{0}) |0\rangle \nn\\
=\, & \langle 0| \ml{O}^\dagger(\vec{0}) e^{iHt_1} n^i_1T_{0i}(r{\vec n}_1) e^{iH(t_2-t_1)} \nn\\
& \qquad\qquad\qquad\qquad \times n^j_2T_{0j}(r{\vec n}_2) e^{-iHt_2} \ml{O}(\vec{0}) |0\rangle \,,
\end{align}
where all the operators in the last two lines are time independent. 

If we can obtain all the eigenvalues and eigenstates of the system, which are defined as $H|m\rangle = E_m|m\rangle$, we can calculate the real-time correlators in \eq{time_correlation1} as
\begin{align}
\label{eq:ee_eigen}
& \langle 0| \ml{O}^\dagger(\vec{0}) n^i_1T_{0i}(t_1,r{\vec n}_1) n^j_2T_{0j}(t_2,r{\vec n}_2) \ml{O}(\vec{0}) |0\rangle \nn\\ 
= \, & e^{i(E_m-E_k)t_1} e^{i(E_k-E_\ell)t_2} \langle 0|\ml{O}^\dagger(\vec{0})|m\rangle \langle m| n^i_1T_{0i}(r{\vec n}_1) | k\rangle \nn\\
&\qquad \qquad \qquad \times \langle k| n^i_2T_{0i}(r{\vec n}_2) | \ell \rangle \langle \ell|\ml{O}(\vec{0})| 0 \rangle \,, 
\end{align}
where sums over the inserted complete sets of states with repeated indices are implicit. In the classical computation approach based on exact diagonalization, which can serve as a useful benchmark for quantum computation on a small lattice, the time integrals can be easily performed
\begin{align}
\label{eqn:phase}
\int_{t_i}^{t_f}\!{\rm d}t\, e^{i\omega t} = \frac{ e^{i\omega t_f} - e^{i\omega t_i}}{i\omega} \,,
\end{align}
where $\omega$ stands for the energy gap between two eigenstates as written in \eq{ee_eigen}. On the other hand, \eq{time_correlation1} will be used for the quantum computation of the energy-energy correlator to be described in the next subsection, as well as for classical computation that is based on sparse matrix multiplication.

\subsection{A quantum algorithm \label{sec:quantum_algorithm}}
\begin{figure}[t!]
\subfloat[Real part.\label{fig:QC_real}]{%
$$\Qcircuit @C=1em @R=1em {
\lstick{\ket{0}} & \qw & \gate{H} & \ctrl{1} & \qw & \gate{H} & \qw &  \meter \\
\lstick{\ket{0}^{\otimes n}} & \gate{U_{\rm grd}} & \qw  & \gate{U} & \qw & \qw & \qw & \qw \\}$$
}
\hfill
\subfloat[Imaginary part.\label{fig:QC_imag}]
{$$\Qcircuit @C=1em @R=1em {
\lstick{\ket{0}} & \qw & \gate{H} & \ctrl{1} & \gate{S^\dagger} & \gate{H} & \qw &  \meter \\
\lstick{\ket{0}^{\otimes n}} & \gate{U_{\rm grd}} & \qw  & \gate{U} & \qw & \qw & \qw & \qw \\ }$$
}
\centering
\caption{Schematic diagrams of the quantum circuits that compute the real and imaginary parts of the two-point real-time correlator $\langle{\rm vac}| \ml{O}^\dagger T_{0i}(t_1) T_{0j}(t_2) \ml{O} |{\rm vac}\rangle $. The system is described by $n$ qubits. A unitary gate $U_{\rm grd}$ is first applied to prepare the ground state, which is followed by the Hadamard test with a controlled-$U$ gate and the top qubit being the ancilla. The $U$ gate implements the $\ml{O}^\dagger T_{0i}(t_1) T_{0j}(t_2) \ml{O}$ operator, which may require additional ancilla qubits themselves that are not shown, since these operators are not unitary. $H$ denotes the Hadamard gate while $S$ represents the phase gate.}
\label{fig:qc_hadamard}
\end{figure}
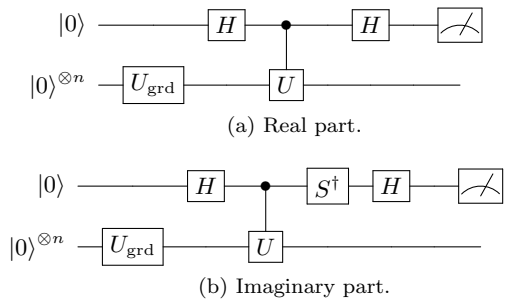

\eq{time_correlation1} informs us on how to construct the quantum algorithm to evaluate the real-time correlator, which is then integrated to obtain the energy-energy correlator as in Eqs.~\eqref{eq:EE_limit1} and~\eqref{eq:lattice_time_inte}. We will now introduce a quantum algorithm that is based on the Hadamard test~\cite{Lin:2022vrd}. A schematic diagram of the algorithm is shown in Fig.~\ref{fig:qc_hadamard}, which consists of the following two steps:
\begin{enumerate}
\item Ground state preparation.

\item Hadamard test with an ancilla qubit. 
\end{enumerate}
We now describe each step in detail.

\subsubsection{Ground state preparation}
The first step of the quantum algorithm prepares the ground state $|{\rm vac}\rangle$ of the interacting system by starting from the qubit default state $|000\dots0\rangle$\footnote{In order to avoid confusion between the interacting vacuum state of the theory $|0\rangle$ and the qubit state $|000\dots0\rangle$, we use $|{\rm vac}\rangle$ to label the former in the description of quantum circuits.} via e.g., the adiabatic time evolution~\cite{Farhi:2000ikn,albash2018adiabatic}. The algorithm simulates the time evolution driven by a time-dependent Hamiltonian
\begin{align}
H_{\rm ad}(t) = H_0 + \frac{t}{T} H_1 \,,
\end{align}
where $H_0$ is a Hamiltonian whose ground state is easy to prepare, such as a product state in the computational basis. $H_0+H_1$ is the target Hamiltonian for which we want to prepare the ground state. The initial state of the time evolution is given by the ground state of $H_0$. The run time $T$ of the adiabatic ground state preparation algorithm scales as $T=O(1/[\min \Delta E(t)]^2)$ where $\Delta E(t) = E_1(t) - E_0(t)$ is the energy gap between the ground state and the first excited state of the time-dependent Hamiltonian $H_{\rm ad}(t)$~\cite{van2001powerful}. For SU(2) and SU(3) non-Abelian gauge theories in two or three spatial dimensions that are confining, we expect a mass gap in the spectrum even in the continuum limit. 

One may also use variational methods, such as the Variational Quantum Eigensolver (VQE) or ADAPT-VQE~\cite{Peruzzo:2013bzg,McClean_2016,PRXQuantum.2.020310,Farrell:2023fgd}, to construct a shallow quantum circuit that prepares the ground state or the state $\ket{\Psi} = \ml{O}|{\rm vac}\rangle$ as well. These algorithms are more empirical than the adiabatic approach in the sense that the variational operator ansatz and their parameters depend on the system details and may need reconstructing at different system sizes and couplings, particularly for the continuum extrapolation.%

In the following, we will use the adiabatic ground state preparation algorithm and leave the exploration of VQE to future work.

\subsubsection{Hadamard test}
After preparing the interacting ground state, we introduce an ancilla qubit and perform the Hadamard test: We first apply a Hadamard gate to an ancilla qubit, then apply a controlled-unitary gate for the operator insertions in the real-time correlator $\langle{\rm vac}| \ml{O}^\dagger T_{0i}(t_1) T_{0j}(t_2) \ml{O} |{\rm vac}\rangle$,\footnote{We suppress the spatial dependence here for convenience. The time dependence is simulated as shown Eq.~\eqref{eq:time_correlation1}.} and finally apply another Hadamard gate and measure the ancilla qubit, as shown in Fig.~\ref{fig:QC_real} The probability of the ancilla qubit being in $|0\rangle$ state is given by
\begin{align}
P(0) = \frac{1+{\rm Re}\langle {\rm vac}| \ml{O}^\dagger T_{0i}(t_1) T_{0j}(t_2) \ml{O} |{\rm vac} \rangle }{2} \,.
\end{align}
When the energy flux operators are spacelike-separated, the energy correlator is real. Thus we do not have to measure its imaginary part. Nevertheless, we discuss how to compute its imaginary part for completeness, which can also serve as a consistency check in practical computations. To measure the imaginary part, we apply a conjugated phase gate $S^\dagger$ to the ancilla qubit right before the second Hadamard gate as shown in Fig.~\ref{fig:QC_imag}. Now the probability of the ancilla qubit being in $|0\rangle$ state is
\begin{align}
P(0) = \frac{1+{\rm Im}\langle {\rm vac}| \ml{O}^\dagger T_{0i}(t_1) T_{0j}(t_2) \ml{O} |{\rm vac} \rangle }{2} \,.
\end{align}

In general, the computational cost of the Hadamard test, i.e., the number of times one applies the $U_{\rm grd}$ and controlled-$U$ gates, scales as $O(1/\epsilon^2)$ for an accuracy $\epsilon$, which can be improved to be $O(1/\epsilon)$ by using the amplitude amplification algorithm, even if the probability of the ``good'' state that we want to amplify is unknown~\cite{Brassard:2000xvp}. However, the amplitude amplification algorithm requires more applications of the $U_{\rm grd}$ and controlled-$U$ gates in the same circuit, resulting in a very deep circuit for the current device. So we will not apply amplitude amplification in our current study.

The source ($\ml{O}$), sink ($\ml{O}^\dagger$) and the energy flux $T_{0i}$ operators are not unitary, making their direct implementation in the quantum circuit hard. Here we assume the source and sink operators to be Hermitian (the energy flux operator $T_{0i}$ is Hermitian) and propose three different methods to implement them. 

\begin{enumerate}[label={\roman*})]
\item Taylor expansion of exponential.
Suppose that $\ml{A}$ is a Hermitian operator and we have to obtain $\ml{A} \ket{\psi}$, we can implement two quantum circuits
where one circuit implements $e^{i \theta \ml{A} }$ gate and the other $e^{-i \theta \ml{A} }$. The difference of the two circuits gives\footnote{In practice, the difference is taken classically on the level of circuit measurement results rather than in the quantum circuit.}
\begin{equation}
\ml{A} \ket{\psi}=  \frac{1}{2 i \theta}\left(e^{i \theta \ml{A} } -e^{-i \theta \ml{A} } \right) \ket{\psi} +O(\theta^2) \,.
\end{equation}
In this way, we can implement $\ml{A} \ket{\psi}$ up to a small error when $\theta$ is small. In our case, we have to implement four different non-unitary operators, $\{\ml{O}^\dagger=\ml{O},\, T_{0i}(t_1),\,T_{0j}(t_2),\,\ml{O} \}$. In order to obtain the real-time correlator $\langle {\rm vac} | \ml{O}  T_{0i}(t_1) T_{0j}(t_2) \ml{O} | {\rm vac} \rangle$, we should run 16 different quantum circuits, in each of which we implement the operator of the form 
$e^{i s_1 \theta \ml{O}} e^{i s_2 \theta T_{0i}(t_1)} e^{i s_3  \theta T_{0j}(t_2)} e^{i s_4 \theta \ml{O}}$ for one sign combination $s_{i} =\pm$. There are a total of 16 sign combinations, corresponding to 16 quantum circuits. By taking a proper linear combination of the results from the 16 quantum circuits, we find 
\begin{equation}
\langle {\rm vac} | \ml{O}  T_{0i}(t_1) T_{0j}(t_2) \ml{O} | {\rm vac} \rangle = \lim_{\theta \rightarrow 0} \frac{M(\theta)}{\theta^4} \,,
\end{equation}
where 
\begin{align}
\qquad M(\theta) &= \frac{1}{16} \sum_{s_1=\pm} \sum_{s_2=\pm} \sum_{s_3=\pm} \sum_{s_4=\pm} (-1)^{s1+s2+s3+s4} \nn\\
& \quad \langle {\rm vac}| e^{i s_1 \theta \ml{O}} e^{i s_2 \theta  T_{0i}(t_1)} e^{i s_3 \theta T_{0j}(t_2)} e^{i s_4 \theta \ml{O} } |{\rm vac} \rangle \nn\\
& = \theta^4 \langle {\rm vac} | \ml{O}  T_{0i}(t_1) T_{0j}(t_2) \ml{O} | {\rm vac} \rangle +O(\theta^6)\,.
\end{align}

The explicit implementation of $e^{i s \theta T_{0i}(t)}$ in practice can be done in three steps: $e^{i s \theta T_{0i}(t)} =e^{i H t} e^{i s \theta T_{0i}(t=0)} e^{-i H t} $, i.e., we perform forward and backward real-time evolution before and after applying the unitary operator $e^{i s \theta T_{0i}(t=0)}$, respectively. Since the signal in $M(\theta)$ is at order $\theta^4$, one needs to achieve at least an accuracy of order $\theta^4$ in each of the 16 circuits.

\item Diluted operator. Similar to the block-encoding method~\cite{Low2019hamiltonian,PhysRevLett.118.010501} and the quantum imaginary time propagation operator~\cite{PhysRevA.105.022440,Turro:2023nuk}, we can add an additional ancilla qubit and implement the following operator\footnote{The normalization of $\ml{A}$ ensures the unitarity of the full operator and increases the success probability.}
\begin{equation}
   U_\ml{A} = \begin{pmatrix}
        \frac{\ml{A}}{\left\|\ml{A}\right\|} & -\sqrt{1-\frac{\ml{A}^2}{\left\|\ml{A}\right\|^2}}\\
        \sqrt{1-\frac{\ml{A}^2}{\left\|\ml{A}\right\|^2}} &  \frac{\ml{A}}{\left\|\ml{A}\right\|}
    \end{pmatrix} \,,
\end{equation}
where the $2\times 2$ block structure is associated with the additional ancilla qubit. After the implementation of $U_\ml{A}$, the ancilla qubit is measured. When the measurement returns $\ket{0}$, the correct operator is applied, i.e., $\ml{A}$. The success probability of applying $\ml{A}$ to $\ket{\psi}$ is given by $P_s=\|  \frac{\ml{A}}{\|\ml{A}\|} |\psi\rangle \|$, which is equal to the expectation value $\langle {\psi} | \frac{\ml{A}^2}{\|\ml{A}\|^2} |{\psi}\rangle$.

\item Linear combination of unitaries (LCU). 
The $\ml{O}$ and $T_{0i}$ operators of our interest can be written as sums of Pauli strings that are unitary operators. Therefore, we can apply the method of LCU~\cite{Childs:2012gwh}.
To implement an operator $\ml{A}$ given by a linear combination of unitary operators $\ml{A} = \sum_{j=1}^N \kappa_j U_j$ with $\kappa_j$ real numbers, we can add $m$ additional ancilla qubits, where $m$ is given by $\lceil \log_2N \rceil$, and implement the quantum circuit shown in Fig.~\ref{fig:LCU_qc}. The $V$ gate is given by a unitary that implements 
\begin{equation}
\label{eqn:LCU_V}
V: |0\rangle^{\otimes m} \to \frac{1}{\sqrt{\| \kappa\|_1}} \sum_{j=1}^N \sqrt{\kappa_j} |j\rangle \,,
\end{equation}
where $\|\kappa\|_1 = \sum_{j=1}^N |\kappa_j|$.
The $U_{\rm LCU}$ gate is constructed as
\begin{equation}
\label{eqn:LCU_U}
U_{\rm LCU} = \sum_{j=1}^N |j\rangle \langle j| \otimes U_j \,.
\end{equation}

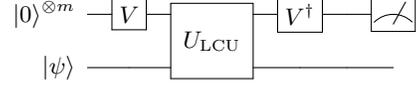
\begin{figure}[t]
\centering
$$\Qcircuit @C=1em @R=1em {
\lstick{\ket{0}^{\otimes m}} & \gate{V} & \multigate{1}{U_{\rm LCU}} &  \gate{V^\dagger} & \qw &  \meter \\
\lstick{\ket{\psi}}  & \qw  & \ghost{U_{\rm LCU}} & \qw & \qw & \qw \\}$$
\caption{LCU quantum circuit. The $V$ and $U_{\rm LCU}$ gates are defined in Eqs.~\eqref{eqn:LCU_V} and~\eqref{eqn:LCU_U}, respectively.}
\label{fig:LCU_qc}
\end{figure}

Analogous to the diluted operator method, the correct implementation of $\ml{A}$ is obtained if all the $m$ ancilla qubits are measured to be $\ket{0}$. The success probability is given by
\begin{equation}
P_s = \frac{ \| \sum_{j=1}^N \kappa_j U_j |\psi\rangle \| }{\| \kappa \|_1^2} \,.
\end{equation}
\end{enumerate}

As we will show later, the $\ml{O}$ and $T_{0i}$ operators of our interest are written as sums of many Pauli strings, which make their implementation on a quantum circuit quite involved via the LCU method.
In this work, we will use the diluted operator method to implement them. 
The real part of the energy-energy corrector is obtained by taking the difference between the probabilities of measuring the ancilla qubit for the Hadamard test to be $\ket{0}$ and $\ket{1}$, respectively, and the other ancilla qubits for the diluted operator method all in $\ket{0}$ states. 

\section{$2+1D$ SU(2) gauge theory on lattice}
\label{sec:su2}
\subsection{Hamiltonian on honeycomb lattice}
We now take the $2+1$-dimensional SU(2) pure gauge theory as an example and perform numerical calculations of the energy correlators on a lattice. Its Lagrangian density takes the form
\begin{align}
\mathcal{L}=-\frac{1}{4g^2} F^{a \mu \nu} F_{\mu \nu}^a\,,
\end{align}
where
\begin{align}
F_{\mu \nu}^a=\partial_\mu A_\nu^a-\partial_\nu A_\mu^a+ \varepsilon^{abc} A_\mu^b A_\nu^c \,,
\end{align}
with $a,b,c \in \{1,2,3\}$ for SU(2) adjoint indices. $\varepsilon^{abc}$ represents the structure constant of the non-Abelian group and is the Levi-Civita tensor for the SU(2) case.

\begin{figure}
\includegraphics[width=0.4\textwidth]{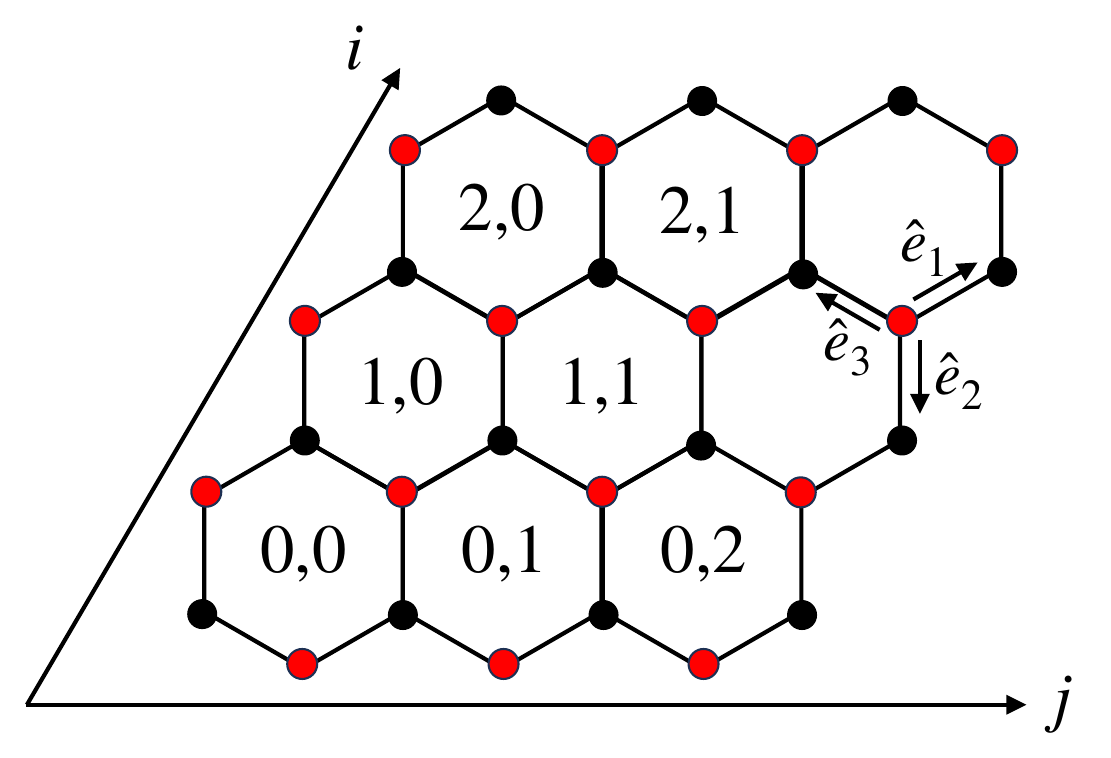}
\caption{A $3\times3$ honeycomb lattice on a plane. The numbers $i,j$ label the central position of each honeycomb cell along the two axes shown, starting from $i=j=0$. They also label the position of the bottom red vertex of each honeycomb cell. The red and black dots represent two types of vertices on the lattice. The three unit vectors $\hat{e}_i$ are defined as starting from red dots and pointing toward black dots. The size of the honeycomb lattice can be specified by the number of plaquettes along the $i$ and $j$ directions, respectively, i.e., $N_i$ and $N_j$.}
\label{fig:honeycomb_lattice}
\end{figure}

We will use the Kogut-Susskind Hamiltonian on a honeycomb lattice shown in Fig.~\ref{fig:honeycomb_lattice}, which can be written as~\cite{Muller:2023nnk}
\begin{align}
\label{eq:HKS}
H &= \frac{3\sqrt{3}g^2}{4}  \sum_{\rm links} E_i^a E_i^a
- \frac{4\sqrt{3}}{9 g^2a^2}  \sum_{\rm plaqs} \varhexagon  \,, \nn\\
\varhexagon &\equiv {\rm Tr}(U_{\varhexagon}) \equiv {\rm Tr} \bigg( \prod_{\substack{{\bs x}\in {\rm red} \\ ({\bs x},\hat{e}_i),({\bs x},\hat{e}_j)\in {\rm plaq} }} U({\bs x},\hat{e}_i) U^\dagger({\bs x},\hat{e}_j) \bigg)\,,
\end{align}
where $a$ in the denominator is the lattice spacing on the honeycomb plane. The coupling constant $g$ has mass dimension of $\frac{1}{2}$ while the electric field operator $E_i^a$ has mass dimension $0$. The index $i$ labels the three unit directions on the honeycomb lattice as shown in Fig.~\ref{fig:honeycomb_lattice} and the electric field is projected along the three directions $E_i^a \equiv \hat{e}_i\cdot {\vec E}^a = \hat{e}_i\cdot (E_x^a, E_y^a)$. We define vertices that move along $\hat{e}_i$ in order to reach neighboring vertices to be red and those along $-\hat{e}_i$ to be black. The magnetic part of the Hamiltonian is given in the second term, which involves the honeycomb plaquette operator $\varhexagon$ that is defined as the trace of the product of the six Wilson lines on the sides of the honeycomb in counter-clockwise direction. Here ${\bs x}$ denotes the red vertex position of one honeycomb plaquette, from which three links begin, but only two lie on the particular honeycomb considered. We label these two directions as $\hat{e}_i$ and $\hat{e}_j$ in the above expression. We use the convention to denote the Wilson line $U$, which can be expressed as an exponential of gauge field, as pointing from red to black vertices. On the other hand, the Wilson line ending on a red vertex is denoted as $U^\dagger$. Further details can be found in Ref.~\cite{Muller:2023nnk}.

Electric fields are generators of time-independent gauge transformation. On the honeycomb lattice, since electric fields live on links, they can generate gauge transformations on either red or black vertices. So we distinguish these two types by the subscript $R$ for red and $B$ for black. The gauge transformation is given by
\begin{align}
\label{eq:gauge_transform}
[E^a_{Bi}({\bs x}+\frac{1}{2}\hat{e}_i), U({\bs x},\hat{e}_j)] &= -\delta_{ij} T^a U({\bs x},\hat{e}_j)  \,, \nn\\
[E^a_{Ri}({\bs x}+\frac{1}{2}\hat{e}_i), U({\bs x},\hat{e}_j)] &= \delta_{ij} U({\bs x},\hat{e}_j) T^a \,,
\end{align}
where $T^a$ denotes the generate of the SU(2) group. The argument of the electric fields, i.e., ${\bs x}+\frac{1}{2}\hat{e}_i$ indicates that they live on links rather than vertices. The gauge transformation generators on the same link satisfy the following commutation relations\footnote{Using the continuum theory, one can derive
\begin{align}
[E^a_{Ri}({\bs x}+\frac{1}{2}\hat{e}_i), U({\bs x},\hat{e}_j)] &= - \delta_{ij} U({\bs x},\hat{e}_j) T^a \,, \nn\\
[E^a_{Ri}, E^b_{Rj}] &= -i\varepsilon^{abc} \delta_{ij} E^c_{Ri} \,. \nn
\end{align}
Flipping the sign of $E_{Ri}^a$ leads to the equations in the main text.}
\begin{align}
\label{eq:[EE]}
[E^a_{Bi}, E^b_{Bj}] &= i\varepsilon^{abc} \delta_{ij} E^c_{Bi} \,, \nn\\
[E^a_{Ri}, E^b_{Rj}] &= i\varepsilon^{abc} \delta_{ij} E^c_{Ri} \,.
\end{align}
Physical states are gauge invariant and thus obey Gauss law 
\begin{align}
\sum_{i=1}^3 E_{i}^a |\psi_{\rm phy}\rangle = 0 \,,
\end{align}
at all vertices (black and red) for all $a$'s (color indices). 

Because of the Peter-Weyl theorem~\cite{Harlow:2018tng}, we can work in the electric basis where the basis states are specified by three quantum numbers $|jm_Rm_B\rangle$, analogous to the total angular momentum quantum number $j$ and two third components $m_R$ and $m_B$ for the red and black vertices, respectively. In the electric basis, the matrix elements of the electric fields are given by~\cite{Byrnes:2005qx,Zohar:2014qma,Liu:2021tef}
\begin{align}
\langle j' m_B' m_R'| E_B^a | j m_B m_R\rangle &= - \delta_{jj'}\delta_{m_Rm_R'} T^{(j)a}_{m_Bm_B'} \,, \nn\\
\langle j' m_B' m_R'| E_R^a | j m_B m_R\rangle &= \delta_{jj'}\delta_{m_Bm_B'} T^{(j)a}_{m_R'm_R} \,,
\end{align}
where $T^{(j)a}$ denotes the generator of the SU(2) irreducible representation with dimension $2j+1$ and is Hermitian: $(T^{(j)a})^* = (T^{(j)a})^T$. The matrix elements of the lattice Wilson line are
\begin{align}
\label{eqn:matrix_U}
&\langle j'm_B'm_R' | U_{s_1s_2} | j m_B m_R \rangle = \nn\\
& \quad \sqrt{\frac{({2j+1})}{({2j'+1})}} \langle j'\, m_B'| j \, m_B; \frac{1}{2}\, s_1 \rangle \langle j\, m_R; \frac{1}{2}\, s_2 | j'\,  m_R' \rangle \,,
\end{align}
where $\langle j'\, m'| j \, m; J\, M \rangle $ denotes the Clebsch-Gordan coefficient.

One can work with only physical states by projecting onto those satisfying Gauss law. After the projection, physical states on the honeycomb lattice can be uniquely determined by just the quantum numbers $j$ on each link and the $m_R$ and $m_B$ quantum numbers can be dropped in the description for any $j_{\rm max}$~\cite{Muller:2023nnk}. This is the advantage of the honeycomb lattice compared with the square lattice, where each vertex has four links, and physical states cannot be uniquely determined in terms of the $j$ values on the four links. This issue of specifying physical states on the square lattice can be overcome by introducing additional labels.

\begin{figure}
\includegraphics[width=0.25\textwidth]{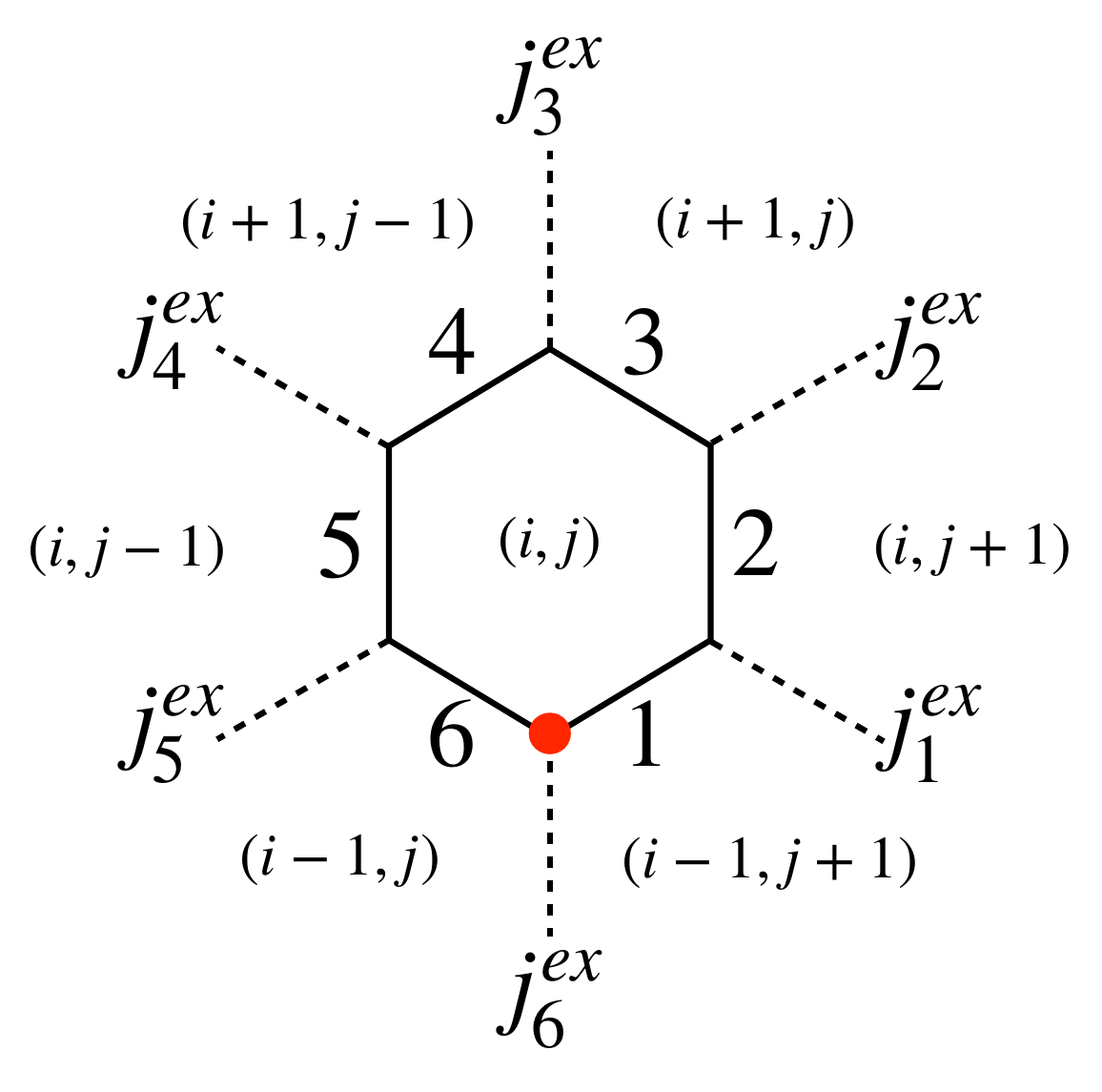}
\caption{A honeycomb with its six internal links (solid lines) labeled by $1,2,3,4,5,6$ and six external links (dashed lines). The electric states on the six external links are specified as shown. $(i,j)$ labels the position of the plaquette under consideration and $(i\pm 1, j\pm1)$ label its six neighbors.}
\label{fig:honeycomb_plaq}
\end{figure}

In the electric basis made up of only physical states, the matrix elements of the electric energy are given by
\begin{align}
\label{eq:elemat}
\langle J | \sum_{a=1}^3(E^a_R)^2| j\rangle = \langle J | \sum_{a=1}^3(E^a_B)^2| j\rangle = j(j+1)\delta_{Jj} \,,
\end{align}
where $|j\rangle$ and $|J\rangle$ denote the states on a link where the electric field squared acts. The matrix elements of the electric energy are diagonal in the electric basis and they are independent of whether the red or black vertex is used. The matrix elements of the magnetic energy (the honeycomb plaquette) are given by~\cite{Zache:2023dko,Hayata:2023puo,Muller:2023nnk} (see Refs.~\cite{Klco:2019evd,ARahman:2021ktn,Hayata:2021kcp,ARahman:2022tkr,Yao:2023pht} for the square plaquette case)
\begin{align}
\label{eq:6j}
&\langle \{J\} | \varhexagon |\{j\} \rangle = \langle J_1J_2\dots J_6 | {\rm Tr}(U_{\varhexagon}) | j_1 j_2\dots j_6 \rangle = \nn\\
&\prod_{i=1}^{6} (-1)^{j_i+J_i+j_i^{ex}} \sqrt{(2J_i+1)(2j_i+1)} \bigg\{ \begin{array}{ccc}  j_i^{ex} & j_i & j_{i+1} \\ \frac{1}{2} & J_{i+1} & J_i   \end{array}  \bigg\} \,,
\end{align}
where we define $j_7=j_1$ and $J_7=J_1$. The curly bracket with six numbers is the Wigner-$6j$ symbol.
The states $|\{j\}\rangle$ and $|\{J\}\rangle$ on both the internal and external links of the honeycomb where the plaquette operator is defined are specified in Fig.~\ref{fig:honeycomb_plaq}. The $j$ values of the external links do not change in the matrix elements.

\subsection{Energy flux operator}
For pure non-Abelian gauge theory, the stress-energy tensor in the continuum is given by
\begin{align}
\label{eqn:Tmunu}
T^{\mu\nu} &= -\frac{1}{g^2} F^{a\mu\rho}F^{a\nu}_{~~~\rho} + \frac{1}{4g^2} \eta^{\mu\nu} F^{a\rho\sigma}F^a_{\rho\sigma} \,.
\end{align}
To construct the energy flow operator, we need specific components of the stress-energy tensor. In particular, the energy \emph{flux} operator is given as
\begin{align}
\label{eq:EFluxcontinuum}
T_{0k}=-T^{0k} = \frac{1}{g^2}F^{a}_{0\ell}F^a_{k\ell} \,,
\end{align}
which we need to integrate in order to build the energy \emph{flow} operator as in~\eq{EF}. Here $k,\ell$ denote spatial indices. Discretized stress-energy tensor operators on square lattices have been constructed in Ref.~\cite{Cohen:2021imf}.

On a honeycomb lattice, the position ${\bs x}$ of the $T_{0k}(\bs{x})$ operator is given by the red vertex as shown in Fig.~\ref{fig:honeycomb_lattice}, which can be labeled by $(i,j)$. The natural directions are no longer $\hat{x}$ and $\hat{y}$, but rather $\hat{e}_1$, $\hat{e}_2$ and $\hat{e}_3$. We can use the lattice discretized version of \eq{EFluxcontinuum} and define three energy flux operators corresponding to the three electric field operators $E_{Ri}^a$ along the three directions $\hat{e}_i$ (see Appendix~\ref{app:T0i}). These energy fluxes point from red vertices to red vertices.
Therefore, to indicate both the spatial position of the energy flux operator and the direction of the energy flow, we find it more convenient to introduce a notation to denote the energy flux from the red vertex $(i,j)$ pointing towards $(i',j')$ as
\begin{align}
\label{eq:Trelabel}
T_{0k}({\bs x}) \to T_{0,(i,j)\to (i',j')}\,.
\end{align}
One can also think of $T_{0,(i,j)\to (i',j')}$ as the energy flux from the plaquette at $(i,j)$ to that at $(i',j')$.

From the expression given in \eq{EFluxcontinuum}, we can express $T_{0,(i,j)\to (i',j')}$ for our $2+1D$ SU(2) on the honeycomb lattice in terms of the electric field and the plaquette operator. We carry out this exercise in Appendix~\ref{app:T0i}. 
However, the energy flux operator obtained by discretizing \eq{EFluxcontinuum} is a lattice bare operator. 
In general, one needs to tame UV divergences and renormalize them in the continuum limit, since the lattice breaks Lorentz symmetry and the stress-energy tensor obtained from discretizing the continuum expression may not satisfy the conservation equation $\partial_\mu T^{0\mu} = 0$ on the lattice. As energy correlators describe how energy flows in the system, it is important to use a lattice energy flux operator that exactly conserves energy in the lattice calculation. Such a physical and IR-finite operator can be constructed by considering the Hamiltonian density at lattice position $(i,j)$ in the Heisenberg picture $H_{ij}(t) = e^{iHt} H_{ij} e^{-iHt}$. Taking time derivative gives
\begin{align}
\frac{{\rm d}H_{ij}(t)}{{\rm d}t} = i[H,H_{ij}(t)] \,.
\end{align}
Comparing with the energy conservation equation
\begin{align}
\partial_t e = -\nabla \cdot {\vec J}^e \,,
\end{align}
where $e$ and ${\vec J}^e$ denotes the energy density and the energy flux respectively, we find the physical and IR-finite energy flux operator from lattice position $(i,j)$ to $(i',j')$ that exactly conserves energy can be written as
\begin{align}
\label{eqn:Jij}
T_{0,(i,j)\to (i',j')} = -i[H_{i'j'}, H_{ij}]\,.
\end{align}
On a honeycomb lattice, six possible energy flux directions $(i',j')$ for a given $(i,j)$ are $(i',j') \in \{(i,j+1),(i+1,j),(i+1,j-1),(i,j-1),(i-1,j),(i-1,j+1)\}$, as shown in Fig.~\ref{fig:honeycomb_plaq}, where the current $T_{0,(i,j)\to (i',j')}$ is nonvanishing. In Appendix~\ref{app:T0i}, we explicitly demonstrate that the lattice bare energy flux operator obtained by discretizing Eq.~\eqref{eq:EFluxcontinuum} is identical to Eq.~\eqref{eqn:Jij} up to boundary terms and UV counterterm. In the following calculations, we will use Eq.~\eqref{eqn:Jij} as the physical and IR-finite energy flux operator.

\subsection{Local Hilbert space truncation at $j_{\rm max}=\frac{1}{2}$}
In practical numerical calculations, one must truncate the local Hilbert space at some value of $j$, namely $j_{\rm max}$. In general, at smaller coupling and/or in order to describe high energy states, one has to use a large value of $j_{\rm max}$. An analytic estimate of the minimum $j_{\rm max}$ needed can be found in Ref.~\cite{Turro:2024pxu}. Here, for simplicity of the demonstration, we just use $j_{\rm max}=\frac{1}{2}$ and leave calculations in the physical limit to future work.

In the case of $j_{\rm max}=\frac{1}{2}$, the Hamiltonian of the $2+1D$ SU(2) lattice gauge theory on a honeycomb lattice can be mapped onto a $2D$ Ising model~\cite{Muller:2023nnk} with the boundary condition that all links outside the honeycomb lattice are in the $j=0$ states. In particular, 
The electric ($H^{\rm el}$) and magnetic ($H^{\rm mag}$) parts of the Hamiltonian can be written as
\begin{align}
\label{eq:H_Ising}
aH &  = H^{\rm el} + H^{\rm mag} \,,  \nn\\
H^{\rm el} & = a \frac{3\sqrt{3}g^2}{4}  \sum_{\rm links} E_i^a E_i^a =h_+ \sum_{(i,j)} \Pi^+_{ij} \nn\\
&\qquad - h_{++} \sum_{(i,j)} \Pi^+_{ij} ( \Pi^+_{i+1,j} + \Pi^+_{i,j+1} + \Pi^+_{i+1,j-1} ) \,, \nn\\
H^{\rm mag} & = - \frac{4\sqrt{3}}{9 g^2a}  \sum_{\rm plaqs} \varhexagon =h_x \! \sum_{(i,j)} \sigma_{ij}^x D_{ij} \,, \nn\\
D_{ij} & = \prod_{K=0}^5 \bigg[ \Big(\frac{1}{2}-\frac{i}{2\sqrt{2}}\Big) \sigma_K^z\sigma_{K+1}^z + \frac{1}{2}+\frac{i}{2\sqrt{2}}\bigg] \,.
\end{align}
Here, the $\Pi^\pm_{ij} = ( 1 \pm \sigma^z_{ij})/2$ operators project onto the spin-up and spin-down states that represent the plaquette state at $(i,j)$ as shown in Fig.~\ref{fig:honeycomb_lattice}. We have multiplied the Hamiltonian by the lattice spacing $a$ such that every quantity has zero mass dimension
\begin{align}
h_+ & = \frac{27\sqrt{3}}{8}ag^2 \,, \quad 
h_{++} = \frac{9\sqrt{3}}{8}ag^2 \,,\quad 
h_x = \frac{4\sqrt{3}}{9ag^2} \,.
\end{align}
The index $K$ stands for a chain $\{K=0: (i,j+1),\ K=1: (i+1,j),\ K=2: (i+1,j-1), \ K=3: (i, j-1), \ K=4: (i-1,j), \ K=5: (i-1, j+1) \}$ and is modulo by six ($K=6$ means $K=0$).

The Hamiltonian density per honeycomb plaquette can be written as
\begin{align}
\label{eqn:Hij}
H_{ij} &= H_{ij}^{\rm el} + H_{ij}^{\rm mag} \nn\\
&= h_+ \Pi^+_{ij} - \frac{h_{++}}{2} \Pi^+_{ij} \sum_{(i',j')}\Pi^+_{i'j'} + h_x \sigma_{ij}^x D_{ij} \,,
\end{align}
where the summation over $(i',j')$ includes $(i,j+1),(i+1,j),(i+1,j-1),(i,j-1),(i-1,j),(i-1,j+1)$. The factor of one half in the second term is inserted to avoid double counting.
Then the energy flux operator can be obtained from Eq.~\eqref{eqn:Jij} 
\begin{align}
\label{eq:energy_energy_flux_operator}
& T_{0,(i,j)\to (i',j')} = -i[H_{i'j'}, H_{ij}] \nn\\
&=  \frac{h_{++}h_x}{2} \big( \Pi^+_{ij} \sigma^y_{i'j'} D_{i'j'} - \Pi^+_{i'j'} \sigma^y_{ij} D_{ij} \big) \nn\\
& + \frac{3h_x^2 }{4} \big( \sigma^x_{i'j'} \sigma^y_{ij} \widetilde{D}_{i'j'} D_{ij} - \sigma^x_{ij} \sigma^y_{i'j'}  \widetilde{D}_{ij} D_{i'j'} \big) \,,
\end{align}
where we have defined
\begin{align}
\widetilde{D}_{i'j'} &\equiv D_{i'j'}\big|_{(\alpha\sigma^z_{i_1j_1}\sigma^z_{ij}+\beta)(\alpha\sigma^z_{ij}\sigma^z_{i_2j_2}+\beta) \to \sigma^z_{i_1j_1} + \sigma^z_{i_2j_2} } \,,\nn\\
\widetilde{D}_{ij} &\equiv D_{ij}\big|_{(\alpha\sigma^z_{i_1'j_1'}\sigma^z_{i'j'}+\beta)(\alpha\sigma^z_{i'j'}\sigma^z_{i_2'j_2'}+\beta) \to \sigma^z_{i_1'j_1'} + \sigma^z_{i_2'j_2'} } \,,
\end{align}
with $\alpha=\frac{1}{2}-\frac{i}{2\sqrt{2}}$ and $\beta=\frac{1}{2}+\frac{i}{2\sqrt{2}}$. The vertical bar means replacement. $D_{ij}$ contains a product over the six nearby indices, among which one is $i'j'$ and is replaced according to the above rule in $\widetilde{D}_{ij}$. Similar replacement is applied in $\widetilde{D}_{i'j'}$. We note that Eq.~\eqref{eq:energy_energy_flux_operator} is the energy flux per plaquette area, rather than per unit area, since the energy density per plaquette area in Eq.~\eqref{eqn:Hij} is used.

For our simulations, we will focus on the energy correlators with local source. For simplicity, we choose both the local source and sink operator to be the magnetic energy term at the central plaquette of the lattice
\begin{align}
\label{eqn:O1/2}
\ml{O} = \ml{O}^\dagger = 
(\sigma_{ij}^x D_{ij})_{\rm center} \,,
\end{align}
where the plaquette position is given by $(i,j)_{\rm center} = (1,1)$ and $(2,2)$ on $3\times 3$ and $5\times 5$ honeycomb lattices, respectively, as shown in Figs.~\ref{fig:honeycomb_lattice} and~\ref{fig:honeycomb_lattice55}. 

\subsection{Quantum circuit for $j_{\rm max}=\frac{1}{2}$}
\label{sec:circuit_1/2} 

We now discuss how to construct the quantum circuit for the $j_{\rm max}=\frac{1}{2}$ case.
To prepare the ground state, we implement the adiabatic evolution starting from the electric ground state that corresponds to $\ket{1}^{\otimes n}$ state,\footnote{In the qubit mapping $|0\rangle$ corresponds to the spin-up state while $|1\rangle$ represents the spin-down state.} where $n=N_iN_j$ is the size of the honeycomb lattice as shown in Fig.~\ref{fig:honeycomb_lattice}.
We apply the Trotter decomposition for adiabatic evolution, which approximates the full evolution operator as
\begin{align}
\ml{T} e^{-i \int_0^{t_{\rm ad}} {\rm d}t' H_{\rm ad}(t') }= \prod_{k=1}^{N_{\rm ad}} \prod_{(i,j)} e^{-i \Delta t H_{ij}^{\rm el}} e^{-i\frac{k}{N_{\rm ad}}\Delta t H_{ij}^{\rm mag} } e^{O(\Delta t)^2} \,,
\end{align}
where $\ml{T}$ is the time-ordering operator and $H_{\rm ad}(t) = H^{\rm el} + \frac{t}{t_{\rm ad}}H^{\rm mag}$ with $t_{\rm ad}=N_{\rm ad}\Delta t_{\rm ad}$.

The implementation of the Hadamard test requires that the $\ml{O}$, $T_{0i}$ and real-time evolution operators are controlled by the ancilla qubit (see Fig.~\ref{fig:qc_hadamard}). A controlled-real-time-evolution operator $C[e^{-iHt}]$ can be written as
\begin{align}
C[e^{-iH t}] & =|0\rangle \langle0| \otimes 1 + |1\rangle \langle1| \otimes e^{-iH t} \nn\\
&= e^{-i \frac{1}{2} (1-\sigma^z) \otimes Ht}\,,
\label{eq:CU}
\end{align}
where $|0\rangle \langle0|$ and $|1\rangle \langle1|$ project the ancilla qubit. 
Moreover, we apply the Trotter decomposition for the real-time evolution
\begin{align}
e^{-i Ht }= \prod_{k=1}^{N_t} \prod_{(i,j)} e^{-i \Delta t H_{ij}^{\rm el}} e^{-i\Delta t H_{ij}^{\rm mag} } e^{O(\Delta t)^2} \,,
\end{align}
where $\Delta t =\frac{t}{N_t}$ and the Hamiltonian terms are given by Eq.~\eqref{eqn:Hij}. Using Eq.~\eqref{eq:CU}, standard circuit compilation for Pauli strings and the Gray code compilation~\cite{Barenco:1995na} for compiling a sum of Pauli-$\sigma^z$ strings, we obtain the quantum circuits for the Trotterized controlled-real-time-evolution operators. 

For the controlled-$T_{0,(i,j)\to (i',j')}$ operators, inspecting Eq.~\eqref{eq:energy_energy_flux_operator}, we notice that the $T_{0,(i,j)\to (i^\prime,j^\prime)}$ operator is written as a sum of six operators ($\sigma^z_{ij} \sigma^y_{i'j'} D_{i'j'}$, $\sigma^y_{i'j'} D_{i'j'}$, $\sigma^x_{i'j'} \sigma^y_{ij} \widetilde{D}_{i'j'} D_{ij}$ and those with $i,j\leftrightarrow i',j'$), which can be implemented in quantum circuits via the diluted operator method. 
The operators $D_{ij}$ and $\tilde{D}_{ij}$ lead to non-unitary evolution, while the rest can be decomposed into $\sigma^x$, $\sigma^y$ and $\sigma^z$ gates and thus can be implemented with standard unitary circuits. 
We compile $D_{ij}$ (and $\widetilde{D}_{ij} $) using the diluted operator method discussed in Sec.~\ref{sec:quantum_algorithm}, by adding an ancilla qubit and implementing the following operator
\begin{align}
U_D = \begin{pmatrix}
\frac{D_T}{\left\|D_T\right\|} & -\sqrt{1-\frac{D_T^2}{\left\|D_T\right\|^2}}\\
\sqrt{1-\frac{D_T^2}{\left\|D_T\right\|^2}} &  \frac{D_T}{\left\|D_T\right\|}
\end{pmatrix}
\end{align}
where $D_T$ is either $D_{ij}$ or $\widetilde{D}_{ij}$. The $U_D$ operator can be written as a unitary evolution operator
\begin{align}
U_D = e^{i\sigma^y\otimes \arccos(\frac{D_T}{\left\|D_T\right\|})}
\end{align}
Similar to Eq.~\eqref{eq:CU}, the controlled-$U_D$ gate can be compiled as
\begin{align}
C[U_D]= e^{i \frac{1}{2} (1-\sigma^z) \otimes \sigma^y \otimes \arccos(\frac{D_T}{\left\|D_T\right\|})} \label{eq:Ut_sigmay}
\end{align}
where the $1-\sigma^z$ acts on the ancilla qubit for the Hadamard test and $\sigma^y$ acts on the ancilla qubit for the $U_D$ implementation.
It is worth noting that $D_T$ is diagonal in the computational basis, which can be decomposed as a sum of Pauli-$\sigma^z$ strings. Furthermore, we can rewrite 
\begin{align}
C[U_D] = S^\dagger H  e^{i\frac{1}{2} (1-\sigma^z) \otimes \sigma^z \otimes \arccos(\frac{D_T}{\left\|D_T\right\|})} H S \,,
\end{align}
where the $S^\dagger H $ and $H S $ gates only act on the ancilla qubit for the $U_D$ implementation, i.e., they are abbreviations of $1\otimes S^\dagger H\otimes 1$ and $1\otimes HS \otimes 1$, respectively.
Applying the Gray code~\cite{Barenco:1995na} for a sum of Pauli-$\sigma^z$ strings, we can compile the $C[U_T]$ operator into quantum circuits. Since each energy flux operator is a sum of six operators that contain $D_{ij}$ ($\widetilde{D}_{ij}$), we have to run 36 quantum circuits in order to compute the energy-energy correlator.

The compilation of the $\ml{O}$ operator follows similarly by using Eqs.~\eqref{eqn:O1/2} and~\eqref{eq:Ut_sigmay}.

\section{Results}
\label{sec:results}
In this section, we present both classical and quantum results of energy correlators on a small honeycomb lattice with $j_{\rm max}=\frac{1}{2}$ and a local source given by Eq.~\eqref{eqn:O1/2}.

\subsection{Classical results}
\subsubsection{$3\times3$ lattice}
\begin{figure}
\includegraphics[width=0.45\textwidth]{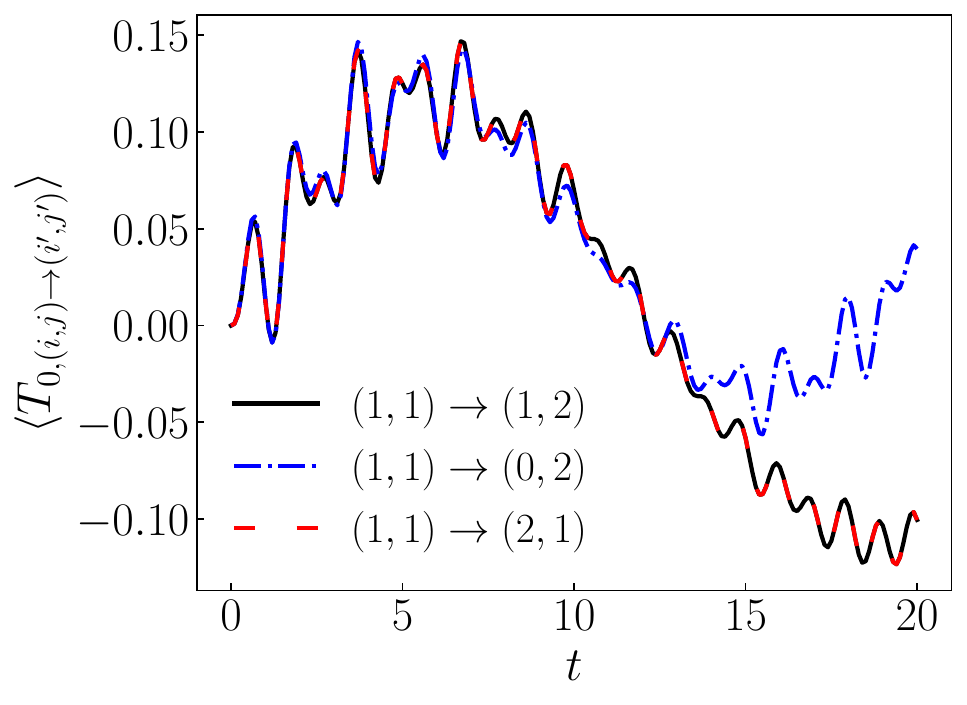}
\caption{Time dependence of energy flux expectation values on a $3\times3$ honeycomb lattice with $ag^2=1$ and $j_{\rm max}=\frac{1}{2}$. The local source term is located at the central plaquette, i.e., $(1,1)$. The energy flux from $(1,1)$ to $(1,2)$ in black and that from $(1,1)$ to $(2,1)$ in red are exactly the same because of the reflection symmetry of the lattice along the axis connecting $(0,0)$ and $(2,2)$. The energy flux from $(1,1)$ to $(0,2)$ in blue agrees with the other two approximately up to $t=13$, which indicates the rotational symmetry is approximately respected at early time. At later time, perturbations reach the boundary and bounce back. The dynamics is influenced by the boundary effect, which breaks the rotational symmetry.}
\label{fig:flux33}
\end{figure}

On a $3\times3$ honeycomb lattice, as shown in Fig.~\ref{fig:honeycomb_lattice}, the local source operator given by the magnetic energy term in Eq.~\eqref{eqn:O1/2} is applied at the central plaquette at position $(1,1)$. We exactly diagonalize the Hamiltonian and calculate energy correlators according to Eqs.~\eqref{eq:ee_eigen} and~\eqref{eqn:phase}. First, as a test, we study the integrand of the one-point energy correlator, i.e., the expectation value of the energy flux operator, as a function of time. Different $\vec{n}$ vectors then pick out different directions of energy flows from the center $(1,1)$. We denote the expectation value of the energy flux operator from $(i,j) \to (i',j')$ at time $t$ as
\begin{align}
\label{eqn:JexpV}
\langle T_{0,(i,j)\to (i',j')} \rangle(t) = \frac{\langle 0| \ml{O}^\dagger e^{iHt} 
 T_{0,(i,j)\to (i',j')} e^{-iHt} \ml{O} | 0\rangle}{\langle 0| \ml{O}^\dagger\ml{O} | 0\rangle} \,,
\end{align}
The results are shown in Fig.~\ref{fig:flux33} for three directions: $(1,1)\to(1,2)$, $(1,1)\to(0,2)$ and $(1,1)\to(2,1)$. The $3\times3$ honeycomb lattice shown in Fig.~\ref{fig:honeycomb_lattice} has two reflection symmetries with respect to the axis connecting $(0,0)$ and $(2,2)$ and that connecting $(0,2)$ and $(2,0)$. Since the local source operator is applied at $(1,1)$ and does not break either reflection symmetry, energy flows in the system still respect both reflection symmetries. As a result, the energy fluxes along $(1,1)\to(1,2)$ and $(1,1)\to(2,1)$ are exactly the same at all time. The system also respects rotational symmetry approximately till the perturbation created by the local source term reaches the boundary and bounces back, which is reflected in the approximate agreement between the energy flux along $(1,1)\to(0,2)$ and the other two energy fluxes in Fig.~\ref{fig:flux33} for $t < 13$. This rotational symmetry is related to the $N=1$ case of \eq{strongcoupl} and shows that the limit of the integral cannot exceed $t = 13$ for one-point correlator.

\begin{figure}
\includegraphics[width=0.43\textwidth]{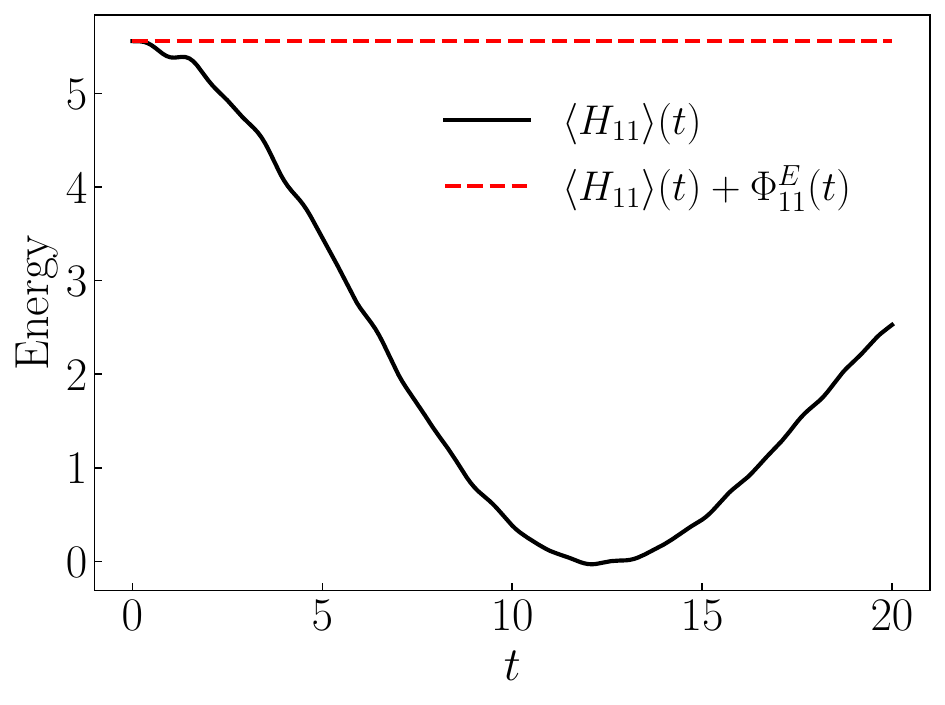}
\caption{Energy of the central plaquette as a function of time (black solid line) compared with the sum of the central plaquette energy and the integrated total energy flux flowing out of the central plaquette (red dashed line) on a $3\times3$ honeycomb lattice with $ag^2=1$ and $j_{\rm max}=\frac{1}{2}$. The constant red dashed line indicates that energy conservation works well in our lattice setup.}
\label{fig:conservation33}
\end{figure}

We further test energy conservation in our lattice setup by computing the energy density per plaquette at position $(1,1)$ and the integrated total energy flux going out of it. The energy density is calculated by taking the expectation value of the Hamiltonian density $H_{ij}$ defined in Eq.~\eqref{eqn:Hij},
\begin{align}
\langle H_{ij} \rangle(t) = \frac{\langle 0| \ml{O}^\dagger e^{iHt} H_{ij} e^{-iHt} \ml{O} | 0\rangle}{\langle 0| \ml{O}^\dagger\ml{O} | 0\rangle} \,.
\end{align}
The integrated total energy flux is calculated by summing over all directions and integrating over time
\begin{align}
\Phi_{ij}^E(t) = \sum_{(i',j')} \int_0^t {\rm d}s\, \langle T_{0,(i,j)\to (i',j')} \rangle(s) \,,
\end{align}
where $i=1$, $j=1$ and the summation over $(i',j')$ includes $(0,1)$, $(0,2)$, $(1,0)$, $(1,2)$, $(2,0)$ and $(2,1)$. The results are depicted in Fig.~\ref{fig:conservation33}. The energy density at $(1,1)$ keeps decreasing till $t\approx 12$, which is roughly the same time when the breaking of the rotational symmetry becomes manifest in Fig.~\ref{fig:flux33}. The energy density at $(1,1)$ starts to increase when the net energy flow bounces back from the boundary and flows inward towards the central plaquette. The energy decrease at position $(1,1)$ is exactly equal to the integrated total energy flux going out of it, as indicated by the constant red dashed line in Fig.~\ref{fig:conservation33}. This demonstrates energy conservation is maintained well in our lattice setup of the energy density and energy flux operators.

\begin{figure}
\includegraphics[width=0.45\textwidth]{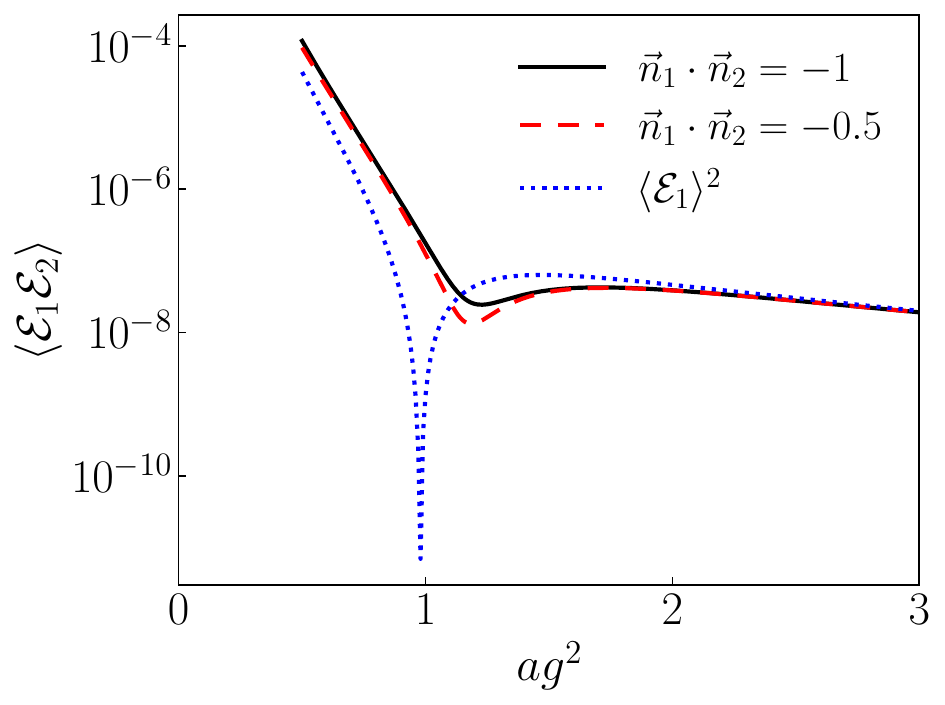}
\caption{Two-point energy correlators as functions of the coupling at two different relative angles: $\theta_{12}=\pi$ (black solid) and $\theta_{12}=\frac{2\pi}{3}$ (red dashed), compared with the one-point energy correlator squared (blue dotted) on a $3\times3$ honeycomb lattice with $j_{\rm max}=\frac{1}{2}$ and a local source at $(1,1)$. The vanishing of the one-point energy correlator around $ag^2=1$ is a lattice artifact, caused by the fact that the energy flux detectors are next to the local source.}
\label{fig:EE33}
\end{figure}

\begin{figure}
\includegraphics[width=0.45\textwidth]{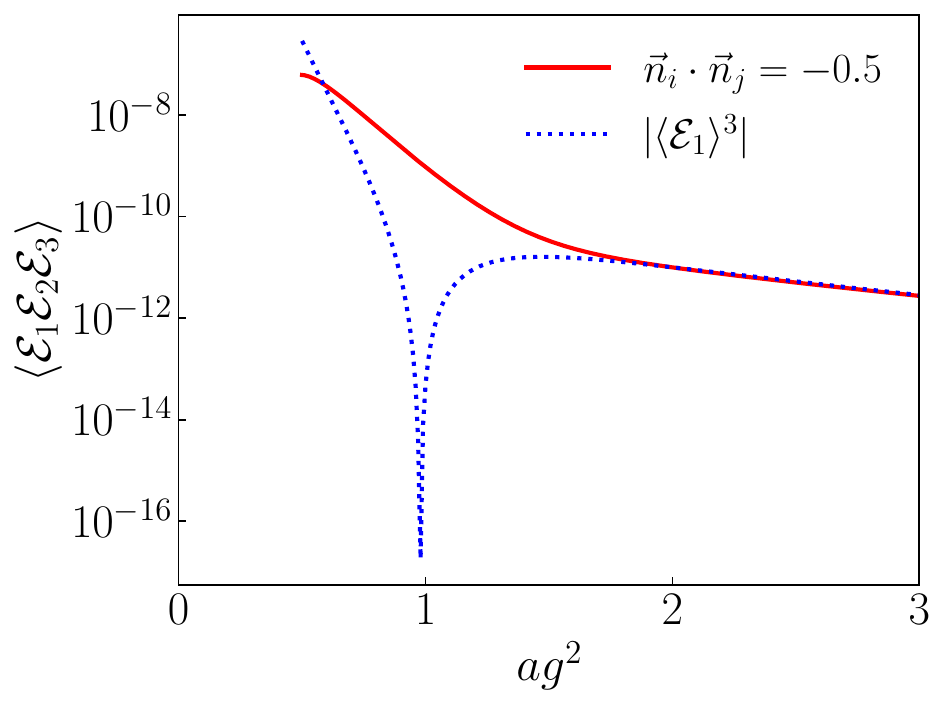}
\caption{Three-point energy correlator as a function of the coupling at $\theta_{12}=\theta_{23}=\theta_{31}=\frac{2\pi}{3}$ (red solid), compared with the magnitude of the one-point energy correlator cubed (blue dotted) on a $3\times3$ honeycomb lattice with $j_{\rm max}=\frac{1}{2}$ and a local source at $(1,1)$. The vanishing of the one-point energy correlator around $ag^2=1$ is a lattice artifact, caused by the fact that the energy flux detectors are next to the local source. As a result, as the coupling decreases the net energy flux starts to flow inward and thus becomes negative in our convention.}
\label{fig:EEE33}
\end{figure}

Finally we study two-point and three-point energy correlators as functions of the coupling. As mentioned in Sec.~\ref{sec:algorithm}, it is important to maintain the spacelike separation between energy flux operators for the whole integration time domain by using the zigzag integration contour. However, on a $3\times 3$ lattice with the source placed at $(1,1)$, we do not have much space to move energy flux operators as time changes. So we just consider the first time integral step in \eq{lattice_time_inte} by fixing the position of the energy flux operator and only integrating for a short period of time, expecting that in such a short time period, the world lines of different energy flux operators will maintain spacelike separation. This requirement immediately excludes the configuration where two energy flux operators are on top of or next to each other.

For two-point energy correlators, we choose $\vec{n}_1$ to be along $(1,1)\to(2,0)$ and we have three available directions for $\vec{n}_2$, which are $(1,1)\to(1,2)$, $(1,1)\to(0,1)$ and $(1,1)\to(0,2)$ respectively. By the reflection symmetry, the directions $(1,1)\to(1,2)$ and $(1,1)\to(0,1)$ are equivalent, leaving us with only two possible nonequivalent directions for $\vec{n}_2$. The results of $\langle \ml{E}_1\ml{E}_2\rangle$ for these two directions are shown in Fig.~\ref{fig:EE33}. We choose the integration time domain to be $t\in[0,0.1]$, which corresponds to choosing $\tilde{t}_1 = 0.1$ for the first step of~\eq{lattice_time_inte}. Slightly changing the upper bound does not change the results qualitatively. Spacelike separation between different energy flux operators can be numerically checked by studying the imaginary part of the correlator. For $\theta_{12}=\pi$, the imaginary part is consistent with zero. For $\theta_{12}=\frac{2\pi}{3}$, the imaginary part is two orders of magnitude smaller than the real part, indicating the spacelike separation is well maintained. On the other hand, if we take $\tilde{t}_1$ to be much larger, then we do observe significant interference between the two detectors that are no longer spacelike separated and results become complex. We also plot the one-point energy correlator squared with the same integration time domain in Fig.~\ref{fig:EE33} for comparison. We find that at strong lattice coupling, the two-point energy correlator factorizes, i.e., $\langle \mathcal{E}_1 \mathcal{E}_2 \rangle \approx \langle \mathcal{E}\rangle^2$. While we find it interesting that our results align with the expectation of the strong coupling regime, we emphasize that one must be careful about interpreting the strength of lattice coupling as the strength of the physical coupling in the theory. In fact, one must take the continuum limit $a\to0$ for physical results, which corresponds to taking the lattice coupling to $0$, i.e., $ag^2\to0$. So the weak or strong coupling in physical limit is not fixed by whether the lattice coupling is small or large, but determined by the energy scale in the physical process calculated. Since we can only study small lattices with the local Hilbert space highly truncated, we will not pursue taking the continuum limit here. Some attempt in dialing the lattice coupling for the continuum limit can be found in Ref.~\cite{Turro:2024pxu}.

We also find that when $ag^2\approx 1$, the one-point energy correlator vanishes. This is a lattice artifact caused by the energy flux operator being next to the local source. At very early time, the energy actually flows inward before flowing outward and this inward-flowing time length increases as the coupling decreases. Since we fix the time integration domain to be $[0,0.1]$, at some coupling, the integrated energy flux vanishes and becomes negative at smaller couplings. As we will see later, on a $5\times5$ lattice we can put the energy flux operator one lattice spacing away from the local source and then this lattice artifact disappears. 

For the three-point energy correlator, we only have one possible configuration that maintains spacelike separation between all three energy flux operators, up to an overall rotation. We choose it to be $\vec{n}_1:(1,1)\to(2,0)$, $\vec{n}_2:(1,1)\to(1,2)$ and $\vec{n}_3:(1,1)\to(0,1)$, which corresponds to $\theta_{ij}=\frac{2\pi}{3}$ between all angles. We use the same time domain for integration, i.e, $t\in[0,0.1]$. The result as a function of coupling is shown in Fig.~\ref{fig:EEE33}, compared with the magnitude of the one-point energy correlator cubed. Again, we observe the factorization of the three-point energy correlator $\langle \ml{E}_1 \ml{E}_2 \ml{E}_3\rangle \approx \langle \ml{E}_1 \rangle^3 $ at strong coupling and the lattice artifact around $ag^2\approx 1$, below which coupling the one-point energy correlator becomes negative, as explained above in the case of two-point energy correlators. 

\subsubsection{$5\times5$ lattice}
\begin{figure}
\includegraphics[width=0.48\textwidth]{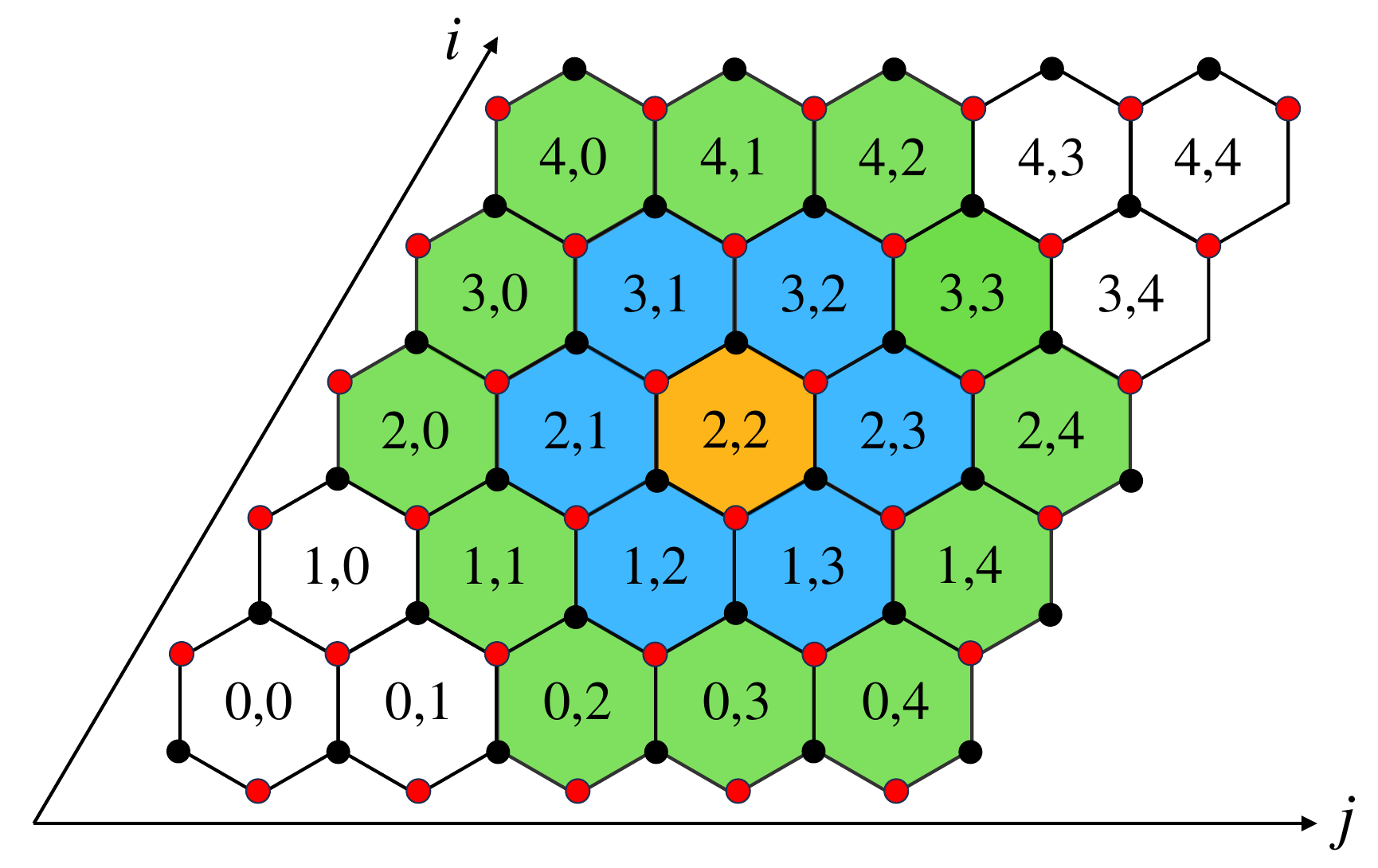}
\caption{A $5\times5$ honeycomb lattice on a plane. The numbers $i,j$ label the central position of each honeycomb cell along the two axes shown, starting from $i=j=0$. The local source is located at the orange plaquette in the center. We consider energy fluxes flowing from the orange plaquette to the blue plaquettes and those flowing from the blue to the green.}
\label{fig:honeycomb_lattice55}
\end{figure}

Next we present classical computing results on a $5\times5$ honeycomb lattice, as depicted in Fig.~\ref{fig:honeycomb_lattice55}. The local source $\ml{O}$ given in Eq.~\eqref{eqn:O1/2} is applied at the central plaquette at $(2,2)$. We consider energy flux operators placed on the blue or green regions of the $5\times5$ honeycomb lattice. The Hilbert space is too large to exactly diagonalize using the computing resources accessible to us. So we use sparse matrix exponentiation and multiplication to compute the real-time correlator in~\eq{time_correlation1} at various time points, and then integrate via a Riemann sum, with a step size ${\rm d}t=0.01$.

\begin{figure}
\subfloat[From orange to blue region.\label{fig:flux55center}]{\includegraphics[width=0.45\textwidth]{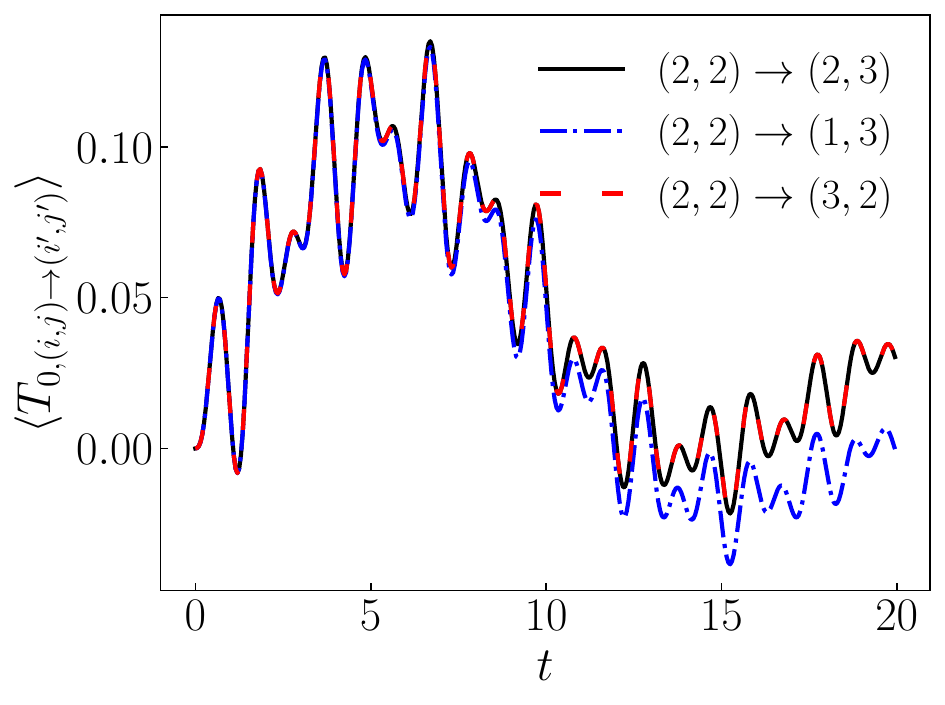}%
}\hfill
\subfloat[From blue to green region.\label{fig:flux55blue}]{\includegraphics[width=0.45\textwidth]{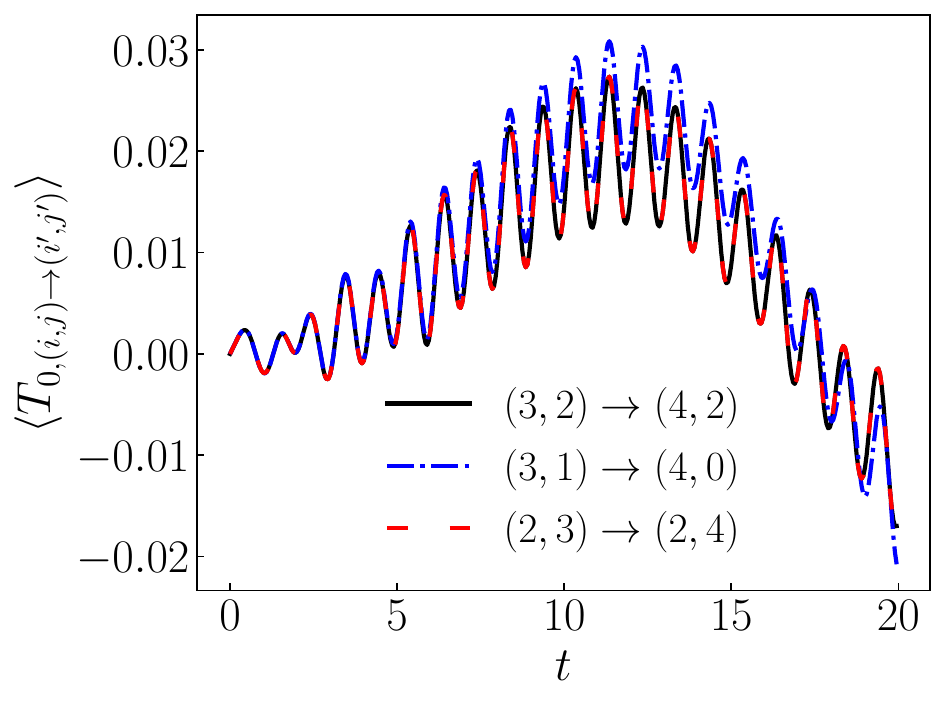}%
}
\caption{Time dependence of energy flux on a $5\times5$ honeycomb lattice with $ag^2=1$ and $j_{\rm max}=\frac{1}{2}$ from orange to blue region (a) and from blue to green region (b). The local source term is applied at the central plaquette, i.e., $(2,2)$. The black solid and red dashed lines in each plot are exactly the same because of the reflection symmetry of the lattice along the axis connecting $(0,0)$ and $(4,4)$. The energy flux marked by the blue dash-dotted line agrees better with the other two in (a) for the time period studied here compared with the $3\times3$ lattice, which indicates the rotational symmetry is better respected on the bigger lattice.}
\label{fig:flux55}
\end{figure}

\begin{figure}
\subfloat[Orange region.\label{fig:55orange}]{\includegraphics[width=0.43\textwidth]{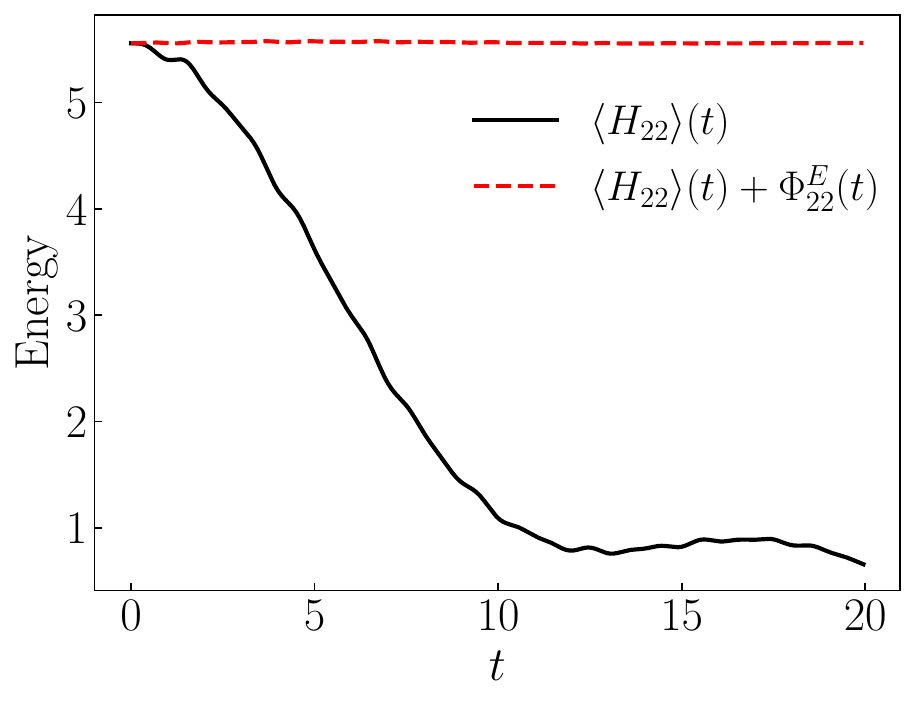}%
}\hfill
\subfloat[Orange and blue regions.\label{fig:55blue}]{\includegraphics[width=0.43\textwidth]{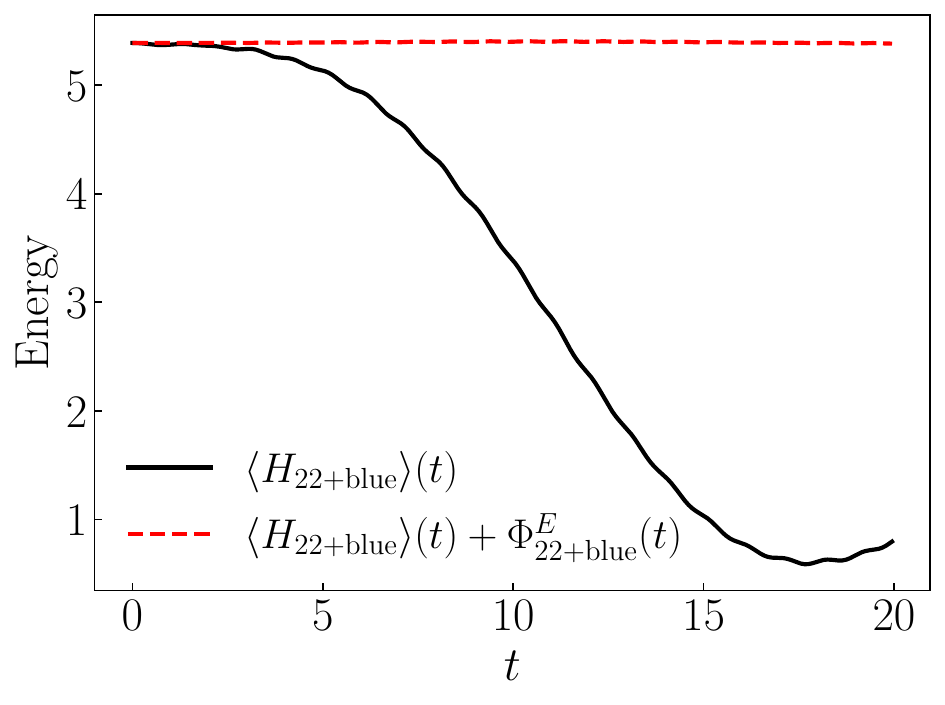}%
}
\caption{Energy conservation test on a $5\times5$ honeycomb lattice with $ag^2=1$ and $j_{\rm max}=\frac{1}{2}$: (a) energy of the central plaquette marked in orange in Fig.~\ref{fig:honeycomb_lattice55} plus the integrated total energy flux going out of the orange plaquette; (b) energies of the plaquettes marked in orange and blue in Fig.~\ref{fig:honeycomb_lattice55} plus the integrated total energy flux going from the blue to the green plaquettes.}
\label{fig:conservation55}
\end{figure}

We first study the time dependence of the expectation value of the energy flux operator as defined in Eq.~\eqref{eqn:JexpV}. We first look at the energy flux from the center into blue plaquettes similar to what we have already studied for the $3\times3$ lattice. In Fig.~\ref{fig:flux55center}, we notice that the boundary effects are significantly reduced compared to the $3\times 3$ case and the rotation symmetry is more respected. In Fig.~\ref{fig:flux55blue}, we also study the energy flux into the green plaquettes, which we can consider for the first time on the $5\times 5$ lattice. As the energy flux originating from the center ends up on green plaquettes only by going through blue plaquettes first, it is sufficient to look at the energy flux from blue to green plaquettes. The energy fluxes along $(3,2)\to(4,2)$ and $(2,3)\to(2,4)$ agree exactly, which is a result of the reflection symmetry of the lattice system along the axis connecting $(0,0)$ and $(4,4)$. The energy flux from $(3,1)$ to $(4,0)$ agrees with the other two approximately for the time period studied here, which indicates the rotational symmetry is approximately respected. The agreement is significantly improved between $t=13$ and $t=20$, compared with the $3\times3$ case in Fig.~\ref{fig:flux33}. We attribute this improvement to the larger lattice with smaller boundary effects.

We further test energy conservation on the $5\times5$ lattice as we did for $3\times 3$ in Fig.~\ref{fig:conservation33}. We perform two tests and present the results in Fig.~\ref{fig:conservation55}. In the first one, we calculate the energy on the central plaquette marked in orange at $(2,2)$ as a function of time and the integrated total energy flux going from the orange plaquette to the blue plaquettes. In the second test, we compare the energies summed over the orange and blue plaquettes and the integrated total energy flux going from the blue to the green plaquettes. In both tests, energy conservation is well maintained. The total energy of the orange and blue plaquettes at $t=0$ is slightly below the energy of the orange plaquette at $t=0$, since the energy density of the ground state is slightly below zero in our convention of the energy reference of the Hamiltonian in Eq.~\eqref{eq:H_Ising}.

\begin{figure}
\includegraphics[width=0.45\textwidth]{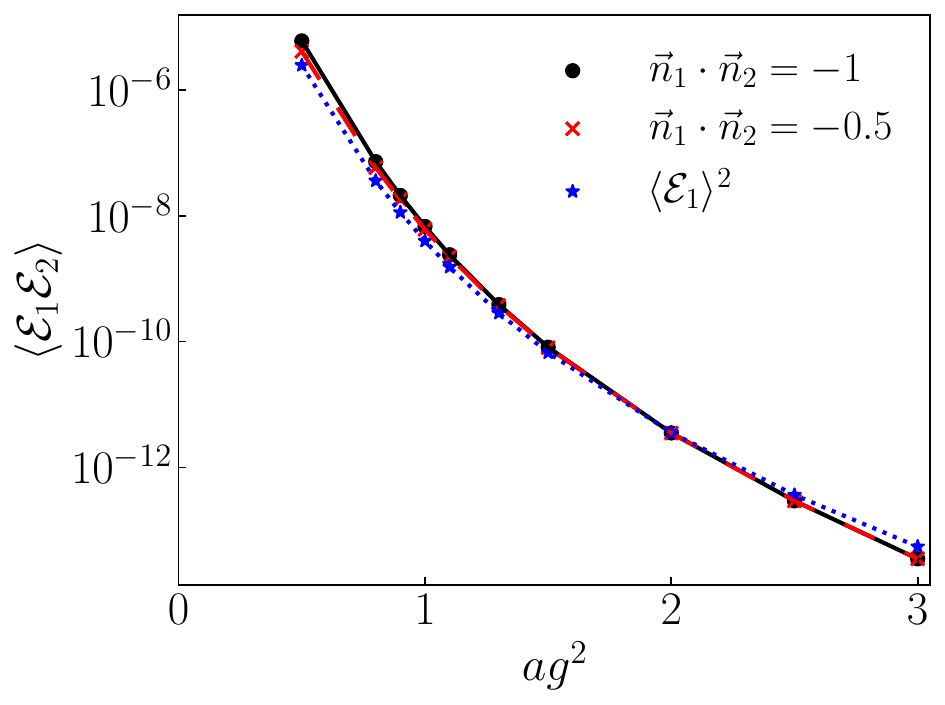}
\caption{Two-point energy correlators as functions of the coupling at two different relative angles: $\theta_{12}=\pi$ (black dots) and $\theta_{12}=\frac{2\pi}{3}$ (red crosses), compared with the one-point energy correlator squared (blue stars) on a $5\times5$ honeycomb lattice with $j_{\rm max}=\frac{1}{2}$ and a local source at $(2,2)$. All energy correlators are calculated from the fluxes going from the blue to green sites, as marked in Fig.~\ref{fig:honeycomb_lattice55}. Since the detectors are separated from the local source by one lattice site, the lattice artifact of vanishing $\langle \mathcal{E}_1\rangle$ around $ag^2=1$ as seen in Figs.~\ref{fig:EE33} and~\ref{fig:EEE33} disappears here.}
\label{fig:EE55green}
\end{figure}

Finally, we calculate two-point energy correlators on the $5\times5$ lattice. In Fig.~\ref{fig:EE55green}, we show the energy-energy correlators with $\theta_{12}=\pi$ and $2\pi/3$ as functions of the coupling, compared with the square of the one-point energy correlator. To compute energy correlators with such angles, we use rotational symmetry and fix specific plaquette sites with such angular separation, rather than compute all the pairs of sites with the same angular separation. In particular, we choose $\vec{n}_1:(3,1)\to(4,0)$ and $\vec{n}_2:(1,3)\to(0,4)$ or $\vec{n}_2:(2,3)\to(2,4)$, i.e., compute the correlation between $(4,0)$ and $(0,4)$ or $(4,0)$ and $(2,4)$. We also compute the one-point correlator along $\vec{n}_1$. We integrate from $t=0$ to $t=0.1$. Approximate factorization is observed for both angles at all the couplings studied here, indicating no jets on our current small and highly truncated lattice. The lattice artifact, i.e., the vanishing of the one-point energy correlator around $ag^2\approx1$ that we observed on the $3\times 3$ lattice in Figs.~\ref{fig:EE33} and~\ref{fig:EEE33} disappears on the $5\times 5$.

\begin{figure*}
\centering
\subfloat[Results for $ag^2=0.6$ with the two detectors at $(0,2)$ and $(2,0)$.]{\includegraphics[width=\columnwidth]{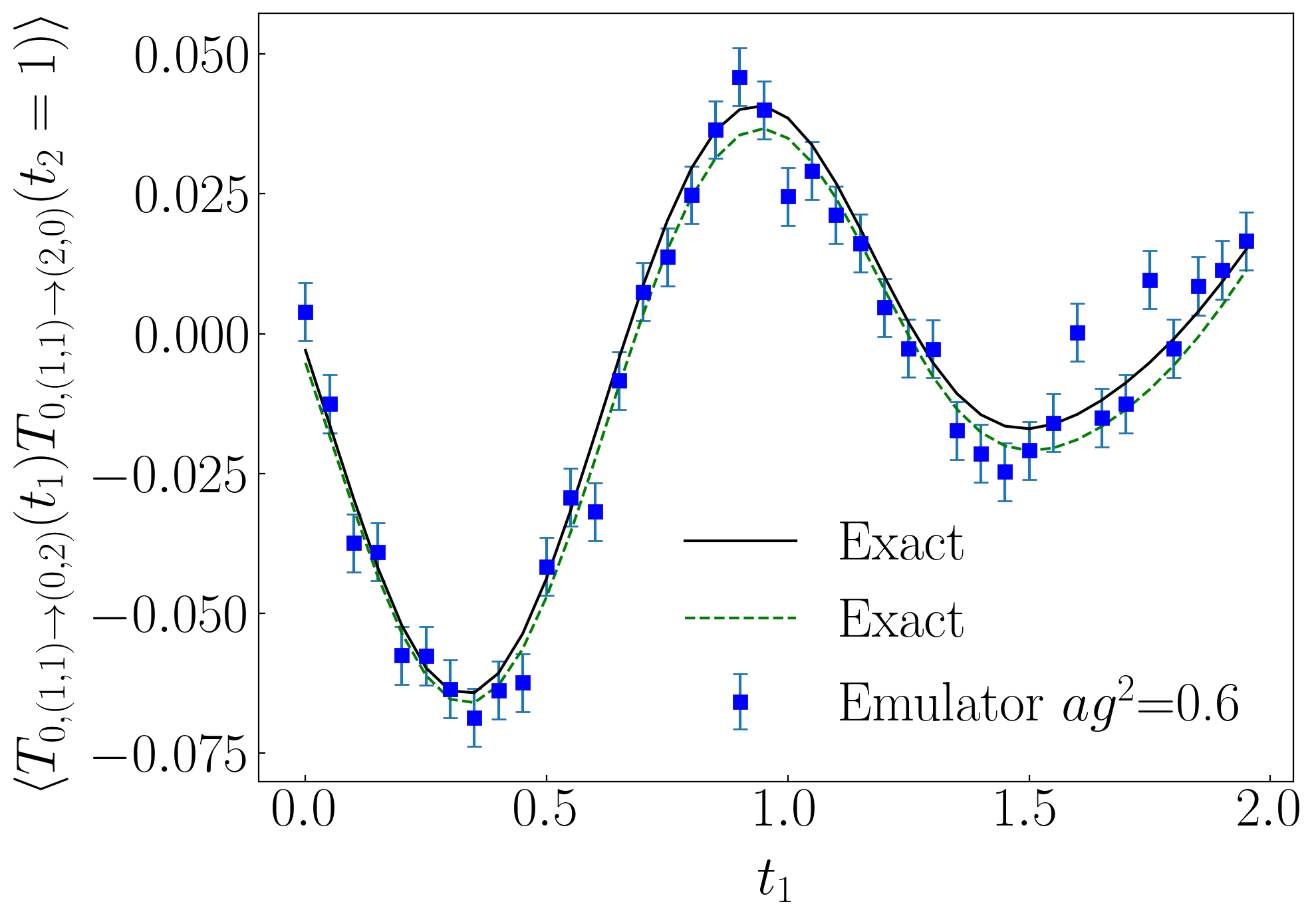}
}\hfill
\subfloat[Results for $ag^2=0.6$ with the two detectors at $(0,2)$ and $(2,1)$.]{\includegraphics[width=\columnwidth]{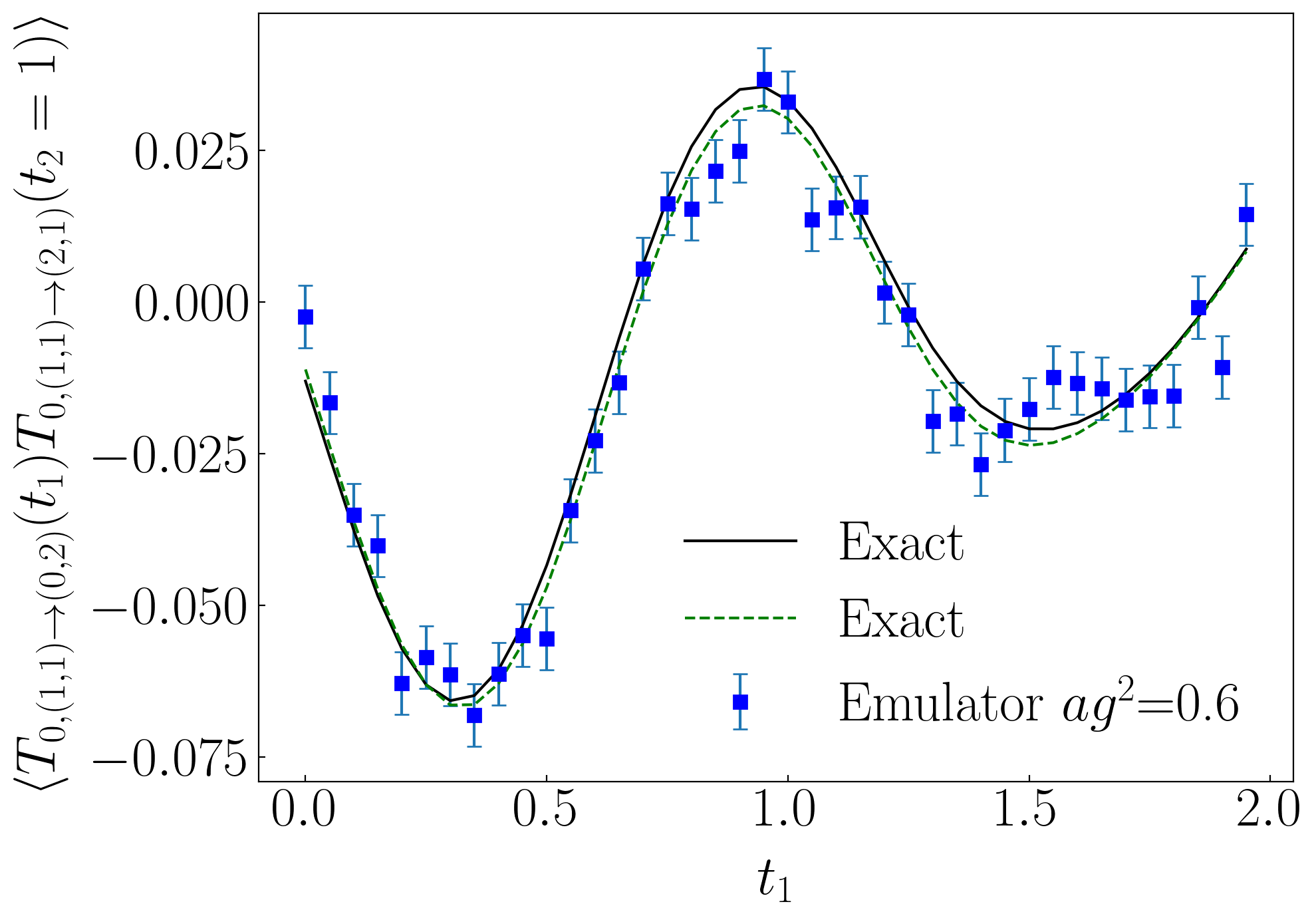}}
 \\
\subfloat[Results for $ag^2=1$ with the two detectors at $(0,2)$ and $(2,0)$.]{\includegraphics[width=\columnwidth]{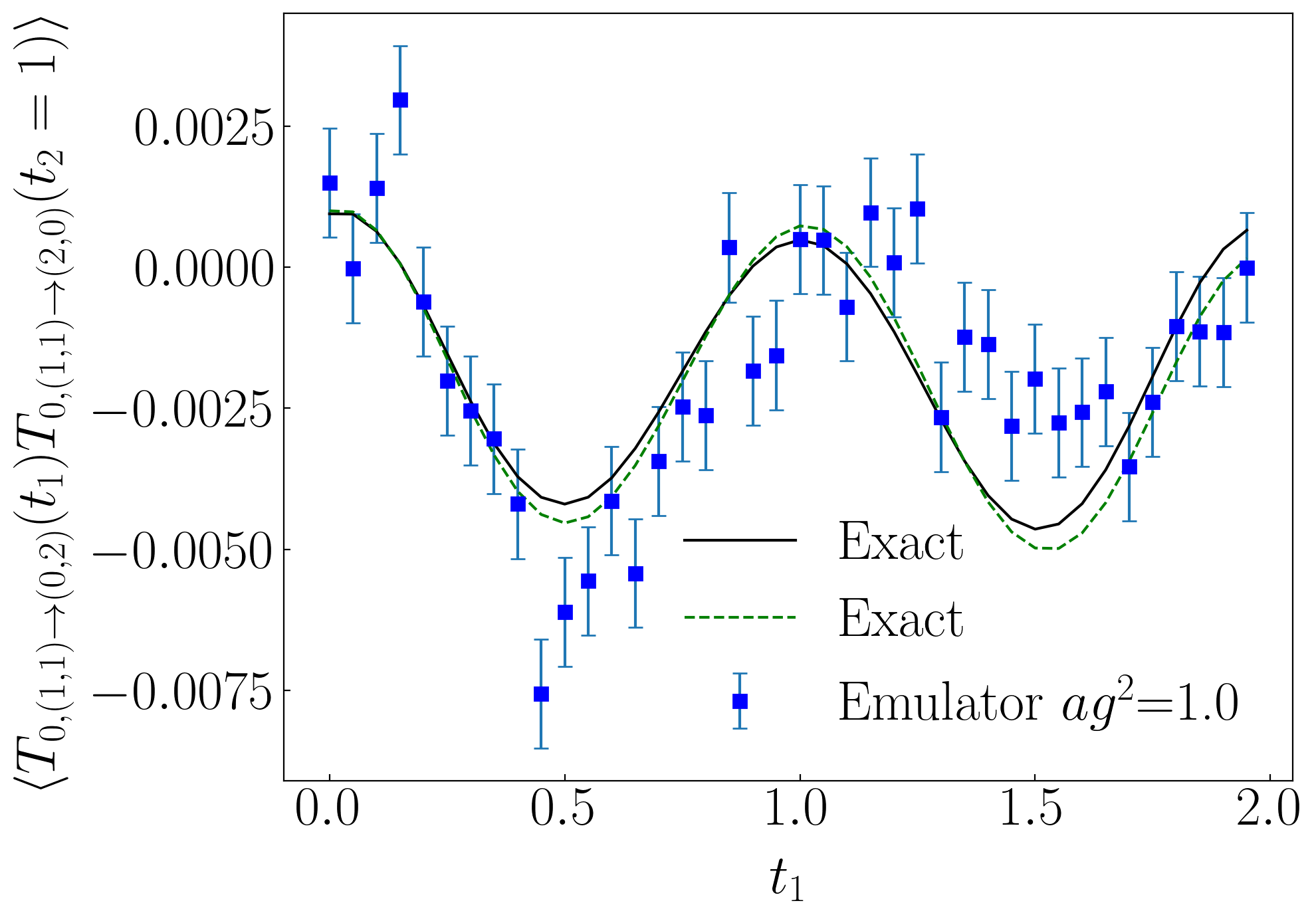}}
\hfill
\subfloat[Results for $ag^2=1$ with the two detectors at $(0,2)$ and $(2,1)$.]{\includegraphics[width=\columnwidth]{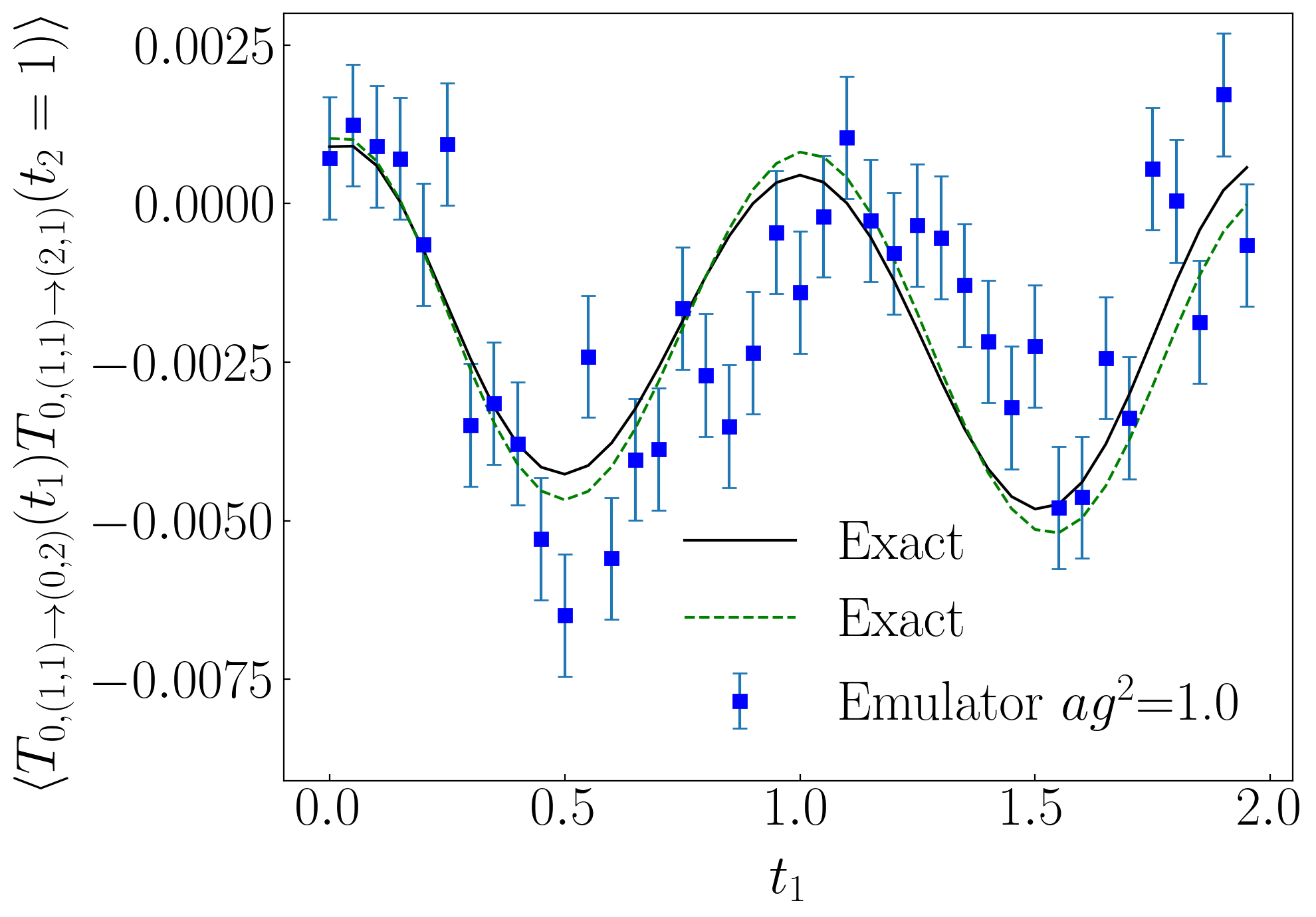}}
\caption{Results from \texttt{IBM Aer} emulator running all 36 quantum circuits for computing the real part of the energy-energy correlator for different detector positions on a $3\times 3$ lattice with $j_{\rm max}=\frac{1}{2}$ at two couplings $ag^2=0.6$ (top row) and $ag^2=1$ (bottom row). The local source is located at plaquette $(1,1)$. Blue points indicate the quantum circuit results obtained from the emulator, the black solid lines label the exact numerical results, and the green dashed lines represent the exact quantum circuit results (including the Trotterization errors in both adiabatic evolution and real-time evolution).  
The energy flux operator is given by Eq.~\eqref{eq:energy_energy_flux_operator}, which is implemented as described in Sec.~\ref{sec:circuit_1/2}.}
\label{fig:energy_energy_qc_3x3}
\end{figure*}

\subsection{Quantum results from Aer emulator}
Finally, we test the reliability of the presented quantum algorithm for computing  energy-energy correlators by using the IBM Aer emulator. In particular, we consider the $3\times3$ honeycomb lattice as shown in Fig.~\ref{fig:honeycomb_lattice} with $j_{\rm max}=\frac{1}{2}$ and two different couplings $ag^2=1$ and $ag^2=0.6$. In the adiabatic evolution, we use $\Delta t_{\rm ad}=0.01$ and $N_{\rm ad}=50$ as the parameters for the $ag^2=1$ case, which gives a fidelity (i.e., the square of the overlap with the true ground state) bigger than $0.995$. For $ag^2=0.6$, using $\Delta t_{\rm ad}=0.01$ and $N_{\rm ad}=300$ leads to a fidelity greater than $0.993$.

We place the two detectors at different locations: the first one at $(i,j)=(0,2)$ and the second one at $(i,j)=(2,1)$ or $(2,0)$. Since our main purpose here is to test the quantum algorithm for calculating the real-time Wightman correlators, rather than obtaining physical results, we will also allow time ranges where the two detectors are no longer spacelike separated. As a result, the energy correlator may become complex. We will only test the quantum algorithms that compute the real part of the energy correlator, though the algorithm to calculate the imaginary part was also described in Fig.~\ref{fig:qc_hadamard}. In particular, we consider $t_1\in[0,2]$ for the first detector with $t_2=1$ fixed for the second detector. In the Trotterization of the real-time evolution, we set $\Delta t=0.025$. As mentioned in Sec.~\ref{sec:circuit_1/2}, we need to run 36 different quantum circuits to calculate the two-point real-time correlator.

Figure~\ref{fig:energy_energy_qc_3x3} shows the results of the energy-energy correlators obtained from the Aer emulator using $2\times10^6$ shots. The top panels show the results for $ag^2=0.6$ while the lower panels for $ag^2=1$. We observe that the results from the quantum emulator (blue squares) give results compatible within two sigmas with the exact numerical results (black solid lines) and the noiseless simulation results (green dashed lines). The errorbars associated with the emulator results are statistical, which are estimated by using binomial distribution and standard uncertainty propagation. The difference between the exact numerical results and the noiseless simulation results originates from the Trotterization in the adiabatic and real-time evolution. We note that in the $ag^2=0.6$ case, the real part of the real-time correlator has a bigger value and thus the statistical uncertainty of using $2\times10^6$ shots is smaller than that in the $ag^2=1$ case using the same number of shots. This indicates that one needs to use more shots for the $ag^2=1$ case in order to achieve the same accuracy as in the $ag^2=0.6$ case.

\section{Conclusions}
\label{sec:conclusions}
In this paper, we present the first study of energy correlators, fundamental Lorentzian field theory observables, using the Hamiltonian lattice approach. We develop a numerical framework for computing energy correlators on the lattice, both for localized sources and for sources with definite momentum, similar to states produced in high-energy collisions. Furthermore, we propose a quantum algorithm based on the Hadamard test to calculate the real-time dependence of Wightman correlation functions, which are essential for computing energy correlators. In implementing the Hadamard test, we need to apply the Hermitian operators for the source, sink, and energy flux in the quantum circuit. We discuss three distinct methods to achieve this: expansion of exponential operators, diluted operator techniques using an ancilla qubit, and linear combinations of unitaries.

To illustrate the utility of our numerical strategy and quantum algorithm, we apply them to the SU(2) pure gauge theory in $2+1$ dimensions discretized on a honeycomb lattice, using the electric basis truncated at $j_{\rm max} = \frac{1}{2}$. With this truncation, the Hamiltonian reduces to a system of interacting spins. We calculate one-, two- and three-point energy correlators on a $3\times 3$ lattice, and one- and two-point correlators on a $5\times 5$ lattice using classical numerical techniques. Our results demonstrate the expected uncorrelated uniform distribution from the strong coupling regime. Finally, we show that our quantum algorithm, when tested on the IBM emulator for the $3\times 3$ lattice, yields results consistent with classical simulations.

Our work marks a significant step in connecting Hamiltonian lattice calculations with real observables in high-energy collisions, while also avoiding the usual challenges of initial wave packet preparation. Our approach provides a complementary pathway to perturbative methods, enabling the exploration of nonperturbative dynamics of energy correlators in confining theories such as QCD, eventually allowing us to probe the confinement transition from partons to hadrons, particularly in light of rapid advances in quantum computing hardware and algorithms. In future work, we plan to improve our lattice calculations by expanding the lattice size and increasing the $j_{\rm max}$ truncation, the latter being crucial for approaching the continuum limit~\cite{Turro:2024pxu}. Additionally, we are interested in extending our studies to gauge theories in $3+1$ dimensions, particularly those involving the SU(3) gauge group and dynamical fermions. Another promising direction is the computation of energy correlators on the Hamiltonian lattice in thermal environments, which has been proposed as a valuable probe for studying quark-gluon plasma in heavy ion collisions~\cite{Bossi:2024qho,Barata:2023bhh,Barata:2023zqg,Andres:2024hdd,Andres:2024ksi,Andres:2022ovj,Yang:2023dwc,Singh:2024vwb}.

\begin{acknowledgements}
We thank Adam Levine, Ian Moult, Natalie Klco, Martin Savage, Matthew Walters, Neill Warrington, and Nikita Zemlevskiy for useful discussions.
K.L. was supported by the U.S. Department of Energy, Office of Science, Office of Nuclear Physics from DE-SC0011090. F.T. and X.Y. were supported by the U.S. Department of Energy, Office of Science, Office of Nuclear Physics, InQubator for Quantum Simulation (IQuS) (https://iqus.uw.edu) under Award Number DOE (NP) Award DE-SC0020970 via the program on Quantum Horizons: QIS Research and Innovation for Nuclear Science.
This research used resources of the National Energy Research Scientific Computing Center (NERSC), a Department of Energy Office of Science User Facility using NERSC award NP-ERCAP0027114.
\end{acknowledgements}

\appendix
\section{Bare stress-energy tensor on the honeycomb lattice}
\label{app:T0i}

On the honeycomb lattice, we can identify the electric field and the plaquette operator as
\begin{align}
\label{eq:honeyEU}
E_x^a &= \frac{a}{g^2}F_{0x}^a \,,\quad E_y^a = \frac{a}{g^2}F_{0y}^a \,, \nn\\
U_{\varhexagon} &= \exp\left( ia^2
\frac{3\sqrt{3}}{2} F_{xy}^aT^a + O(a^3)\right)\,.
\end{align}
Using this identification naively, we can obtain the bare lattice version of the energy flux operator defined in \eq{EFluxcontinuum}. Its $x$ and $y$ spatial components are given as
\begin{align}
\label{eqn:T0i_lattice}
T^{\rm bare}_{0x} &= \frac{-i}{3\sqrt{3}a^3} \big( E_y^a {\rm Tr}( T^a U_{\varhexagon}) + {\rm Tr}( T^a U_{\varhexagon}) E_y^a  \nn\\
&\qquad\qquad - E_y^a {\rm Tr}( T^a U_{\varhexagon}^\dagger) - {\rm Tr}( T^a U_{\varhexagon}^\dagger) E_y^a \big) \,,\nn\\
T^{\rm bare}_{0y} &= \frac{i}{3\sqrt{3}a^3} \big( E_x^a {\rm Tr}( T^a U_{\varhexagon}) + {\rm Tr}( T^a U_{\varhexagon}) E_x^a  \nn\\
&\qquad\qquad  - E_x^a {\rm Tr}( T^a U_{\varhexagon}^\dagger) - {\rm Tr}( T^a U_{\varhexagon}^\dagger) E_x^a \big) \,,
\end{align}
where we have used the difference between $U_{\varhexagon}$ and $U_{\varhexagon}^\dagger$ to reduce the lattice discretization error. Different orderings of the operators $E^a_i$ and ${\rm Tr}(T^a U_{\varhexagon})$ may give different matrix elements on the lattice, so we take the average of the two orderings. 

However, Eq.~\eqref{eqn:T0i_lattice} so far has not fully accounted for the honeycomb lattice structure. First, we note that the energy flux in Eq.~\eqref{eqn:T0i_lattice} is a density per unit area $a^2$. In the main text we study energy density and energy flux density per honeycomb plaquette. So to match with the main text, we need to multiply a factor of the honeycomb plaquette area, which is $\frac{3\sqrt{3}}{2}a^2$, i.e., 
\begin{align}
\label{eqn:T0i_area}
T^{\rm bare}_{0i} \to \frac{3\sqrt{3}}{2}a^2 T^{\rm bare}_{0i} \,.
\end{align}
Second, $\hat{x}$ and $\hat{y}$ are not the natural directions of the honeycomb lattice. As shown in Fig.~\ref{fig:honeycomb_lattice}, the three natural directions for electric fields on the links are denoted as $\hat{e}_i$ for $i=1,2,3$. We now try to construct energy flux operators using the directions more natural to the honeycomb lattice.

We note that the above expression of $T^{\rm bare}_{0x}$ contains only $E_y^a$ while that of $T^{\rm bare}_{0y}$ contains only $E_x^a$, indicating that the direction of the energy flux operator is perpendicular to that of the electric field, as in the Poynting vector. %
This observation informs us how to generalize Eq.~\eqref{eqn:T0i_lattice} for the honeycomb lattice: the expression containing the electric field on a given link describes the energy flux across that link.

For a plaquette located at position $(i,j)$ as in Fig.~\ref{fig:honeycomb_plaq}, the six possible energy flux directions $(i',j')$ are given as $(i',j') \in \{(i-1,j+1),(i,j+1),(i+1,j),(i+1,j-1),(i,j-1),(i-1,j)\}$, corresponding directly to the energy flux across the six links labeled 1 through 6, respectively. As discussed around \eq{Trelabel}, we can label associated energy flux operator between $(i,j)$ and $(i',j')$ as $T^{\rm bare}_{0,(i,j)\to (i',j')}$. For example, if we consider electric fields pointing from the red vertex along the link 1 and link 6, we find these electric fields are related to the energy fluxes between $(i,j) \to (i-1,j+1)$ and $(i,j) \to (i-1,j)$, respectively. As shown in Fig.~\ref{fig:honeycomb_lattice}, these two directions on link 1 and 6 are denoted as $\hat{e}_1$ and $\hat{e}_3$, and thus the electric fields along them starting from the red vertex are denoted as $E_{R1}^a$ and $E_{R3}^a$, respectively. As our plaquette operator $U_{\varhexagon}$ is defined with the convention that traces over six gauge links in counter-clockwise direction, it is useful to consider both electric fields pointing from the red vertex along the link 1 in $\hat{e}_1$ direction (counter-clockwise direction) and along the link 6 in $\hat{e}_3$ direction (clockwise direction).

Shifting the $E_x^a$ or $E_y^a$ from Eq.~\eqref{eqn:T0i_lattice} to $E_{R1}^a$ and $E_{R3}^a$, which respectively indicates the electric field pointing along the link 1 and 6 from the red vertex, the relevant matrix elements of the energy flux operator for link 1 in the physical Hilbert space is given by
\begin{align}
& \langle \{ J \} | {\rm Tr}(T^a U_{\varhexagon})E_{R1}^a | \{ j \} \rangle = - \frac{\langle \{ J \} | {\rm Tr}(U_{\varhexagon}) | \{ j \} \rangle }{\begin{Bmatrix} 
j_6^{ex} & j_6 & j_1 \\
\frac{1}{2} & J_1 & J_6 
\end{Bmatrix} }  \nn\\
&\times (-1)^{s_1'+j_6^{ex}-M_{6}-M_{1}+J_1-j_1} T^a_{s_1s_1'} T^{(j_1)a}_{m_{1}m_{1}'}
{\begin{pmatrix} 
j_6 & j_6^{ex} & j_1 \\
m_{6} & m_6^{ex} & m_{1} 
\end{pmatrix} }  \nn\\
&\times
{\begin{pmatrix} 
J_6 & j_6^{ex} & J_1 \\
M_{6} & m_6^{ex} & M_{1} 
\end{pmatrix} }
{\begin{pmatrix} 
j_1 & \frac{1}{2} & J_1 \\
m_{1}' & s_1 & -M_{1} 
\end{pmatrix} }
{\begin{pmatrix} 
j_6 & \frac{1}{2} & J_6 \\
m_{6} & -s_1' & -M_{6} 
\end{pmatrix} } \,, \nn\\
& \langle \{ J \} | E_{R1}^a{\rm Tr}(T^a U_{\varhexagon}) | \{ j \} \rangle = - \frac{\langle \{ J \} | {\rm Tr}(U_{\varhexagon}) | \{ j \} \rangle }{\begin{Bmatrix} 
j_6^{ex} & j_6 & j_1 \\
\frac{1}{2} & J_1 & J_6 
\end{Bmatrix} }  \nn\\
&\times (-1)^{s_1'+j_6^{ex}-M_{6}-M_{1}'+J_1-j_1} T^a_{s_1s_1'} T^{(J_1)a}_{M_{1}'M_{1}}
{\begin{pmatrix} 
j_6 & j_6^{ex} & j_1 \\
m_{6} & m_6^{ex} & m_{1} 
\end{pmatrix} }  \nn\\
&\times
{\begin{pmatrix} 
J_6 & j_6^{ex} & J_1 \\
M_{6} & m_6^{ex} & M_{1} 
\end{pmatrix} }
{\begin{pmatrix} 
j_1 & \frac{1}{2} & J_1 \\
m_{1} & s_1 & -M_{1}' 
\end{pmatrix} }
{\begin{pmatrix} 
j_6 & \frac{1}{2} & J_6 \\
m_{6} & -s_1' & -M_{6} 
\end{pmatrix} } \,,
\end{align}
where repeated indices are summed over. The internal and external link states are specified as in Fig.~\ref{fig:honeycomb_plaq}. The big parenthesis with six numbers is the Wigner-$3j$ symbol. 
Similarly, the expression of the energy flux associated with the electric field along link 6 from the red vertex is given by
\begin{align}
& \langle \{ J \} | {\rm Tr}(T^a U_{\varhexagon})E_{R3}^a | \{ j \} \rangle = - \frac{\langle \{ J \} | {\rm Tr}(U_{\varhexagon}) | \{ j \} \rangle }{\begin{Bmatrix} 
j_6^{ex} & j_6 & j_1 \\
\frac{1}{2} & J_1 & J_6 
\end{Bmatrix} }  \nn\\
&\times (-1)^{s_1'+j_6^{ex}-M_{6}-M_{1}+J_1-j_1} T^a_{s_1s_1'} T^{(j_6)a}_{m_{6}m_{6}'}
{\begin{pmatrix} 
j_6 & j_6^{ex} & j_1 \\
m_{6} & m_6^{ex} & m_{1} 
\end{pmatrix} }  \nn\\
&\times
{\begin{pmatrix} 
J_6 & j_6^{ex} & J_1 \\
M_{6} & m_6^{ex} & M_{1} 
\end{pmatrix} }
{\begin{pmatrix} 
j_1 & \frac{1}{2} & J_1 \\
m_{1} & s_1 & -M_{1} 
\end{pmatrix} }
{\begin{pmatrix} 
j_6 & \frac{1}{2} & J_6 \\
m_{6}' & -s_1' & -M_{6} 
\end{pmatrix} } \,, \nn\\
& \langle \{ J \} | E_{R3}^a{\rm Tr}(T^a U_{\varhexagon}) | \{ j \} \rangle = - \frac{\langle \{ J \} | {\rm Tr}(U_{\varhexagon}) | \{ j \} \rangle }{\begin{Bmatrix} 
j_6^{ex} & j_6 & j_1 \\
\frac{1}{2} & J_1 & J_6 
\end{Bmatrix} }  \nn\\
&\times (-1)^{s_1'+j_6^{ex}-M_{6}'-M_{1}+J_1-j_1} T^a_{s_1s_1'} T^{(J_6)a}_{M_{6}'M_{6}}
{\begin{pmatrix} 
j_6 & j_6^{ex} & j_1 \\
m_{6} & m_6^{ex} & m_{1} 
\end{pmatrix} }  \nn\\
&\times
{\begin{pmatrix} 
J_6 & j_6^{ex} & J_1 \\
M_{6} & m_6^{ex} & M_{1} 
\end{pmatrix} }
{\begin{pmatrix} 
j_1 & \frac{1}{2} & J_1 \\
m_{1} & s_1 & -M_{1} 
\end{pmatrix} }
{\begin{pmatrix} 
j_6 & \frac{1}{2} & J_6 \\
m_{6} & -s_1' & -M_{6}' 
\end{pmatrix} } \,.
\end{align}

These expressions are valid for general values of $j_{\rm max}$. If we now take the case of $j_{\rm max}=\frac{1}{2}$, then we can use the matrix elements of the magnetic energy term (the plaquette operator) in \eq{6j} to further simplify these in terms of Pauli matrices
\begin{align}
({\rm Tr}(T^a U_{\varhexagon})E_{R1}^a)_{ij} &= - \frac{3}{8}\varhexagon_{ij} (1-\sigma^z_{ij}\sigma^z_{i-1,j+1}) \,,\nn\\
(E_{R1}^a{\rm Tr}(T^a U_{\varhexagon}))_{ij} &= \frac{3}{8} (1-\sigma^z_{ij}\sigma^z_{i-1,j+1}) \varhexagon_{ij} \,. \nn\\
({\rm Tr}(T^a U_{\varhexagon})E_{R3}^a)_{ij} &= \frac{3}{8}\varhexagon_{ij} (1-\sigma^z_{ij}\sigma^z_{i-1,j}) \,,\nn\\
(E_{R3}^a{\rm Tr}(T^a U_{\varhexagon}))_{ij} &= -\frac{3}{8} (1-\sigma^z_{ij}\sigma^z_{i-1,j}) \varhexagon_{ij} \,.
\end{align}
More specifically, using $\varhexagon_{ij} = \sigma_{ij}^xD_{ij}$ as in~\eq{H_Ising}, we find
\begin{align}
\label{eqn:T0i_jmax1/2}
\{{\rm Tr}(T^a U_{\varhexagon}), E_{R1}^a\}_{ij} &=  -\frac{3i}{4} \sigma_{ij}^y \sigma^z_{i-1,j+1}  D_{ij} \,,\nn\\
\{{\rm Tr}(T^a U_{\varhexagon}),E_{R3}^a\}_{ij} & = \frac{3i}{4} \sigma_{ij}^y \sigma^z_{i-1,j} D_{ij} \,.
\end{align}
It is important to note that the two operators in Eq.~\eqref{eqn:T0i_jmax1/2} are anti-Hermitian, which guarantees that the energy flux operator $T^{\rm bare}_{0i}$ is Hermitian due to the factor of $i$ in Eq.~\eqref{eqn:T0i_lattice} and thus is a valid observable.

Finally, we note that we also need terms involving $U_{\varhexagon}^\dagger$ as indicated in Eq.~\eqref{eqn:T0i_lattice}. In the plaquette at $(i,j)$ shown in Fig.~\ref{fig:honeycomb_plaq}, the Wilson line on link 1 inside $U_{\varhexagon}$ starts on the red dot, i.e., it is $U_1$. The $U_{\varhexagon}^\dagger$ as needed in Eq.~\eqref{eqn:T0i_lattice} is obtained from the plaquette at $(i-1,j+1)$ since the Wilson line on link 1 there ends on the red dot, i.e., it is $U_1^\dagger$. A similar argument applies to link 6. Putting all these together, we find the energy flux from plaquette $(i,j)$ to plaquette $(i-1,j+1)$ or $(i-1,j)$ is given by per plaquette area in lattice units
\begin{align}
&T^{\rm bare}_{0,(i,j)\to (i-1,j+1)} \nn\\
&\quad = \frac{3}{8} (\sigma_{i-1,j+1}^y \sigma^z_{ij}  D_{i-1,j+1} - \sigma_{ij}^y \sigma^z_{i-1,j+1}  D_{ij} ) \,,\nn\\
&T^{\rm bare}_{0,(i,j)\to (i-1,j)} \nn\\
&\quad = \frac{3}{8} (\sigma_{i-1,j}^y \sigma^z_{ij}  D_{i-1,j} - \sigma_{ij}^y \sigma^z_{i-1,j}  D_{ij} ) \,.
\end{align}
We can work out the similar formula for $T_{0,(i,j)\to (i',j')}$ for the other four neighboring plaquettes as well. This can be generally casted into the form 
\begin{align}
\label{eq:energy_flux_operator_wobd}
& T^{\rm bare}_{0,(i,j)\to (i',j')} = \frac{3}{8} \big( \sigma^y_{i'j'} \sigma^z_{i j} D_{i'j'} - \sigma^y_{ij} \sigma^z_{i'j'} D_{ij} \big) \,.
\end{align}

This expression differs slightly from the IR-finite one derived in \eq{energy_energy_flux_operator}. It is the same as \eq{energy_energy_flux_operator} up to a lattice boundary effect and UV counterterm. First, the extra term in the second line of \eq{energy_energy_flux_operator} only exists for the boundary plaquette, which can be ignored in the large lattice size limit. The remaining difference is the last line of \eq{energy_energy_flux_operator}, which diverges as $1/(a^2g^4)$ in the continuum limit and can be thought of as UV counterterm.

\bibliography{main.bib}
\bibliographystyle{apsrev4-1}

\end{document}